\documentclass[12pt]{article}
\usepackage[dvipsnames]{xcolor}
\usepackage{titlesec}
\usepackage{multirow}
\usepackage{amsmath,amsfonts,amssymb} 

\usepackage{enumerate}
\usepackage{url} 
\usepackage{bbm}
\usepackage{lscape}
\usepackage{bm}
\usepackage[font=small,labelfont=bf]{caption}
\usepackage{graphicx}
\usepackage{dsfont}
\usepackage[normalem]{ulem}

\usepackage{wrapfig}
\usepackage{algorithm}
\usepackage{subcaption}
\usepackage{soul}

\usepackage{titletoc}

\newcommand\DoToC{%
\startcontents[appendices]
\printcontents[appendices]{l}{1}{\section*{Appendices}\setcounter{tocdepth}{2}}
}

\usepackage{amsthm} 

\usepackage{tabularx}
\usepackage[table]{}

\usepackage{threeparttable}
\usepackage{dcolumn, booktabs}
\usepackage{lipsum}

\usepackage{ctable}

\usepackage{array}
\newcommand{\PreserveBackslash}[1]{\let\temp=\\#1\let\\=\temp}
\newcolumntype{C}[1]{>{\PreserveBackslash\centering}p{#1}}
\newcolumntype{R}[1]{>{\PreserveBackslash\raggedleft}p{#1}}
\newcolumntype{L}[1]{>{\PreserveBackslash\raggedright}p{#1}}

\usepackage[T1]{fontenc}
\usepackage{mathptmx}
\usepackage{mathtools}

\usepackage{setspace}
\usepackage{arydshln}

\usepackage{hyperref}
\hypersetup{
    colorlinks=true,
    linkcolor=blue,
    filecolor=blue,      
    urlcolor=blue,
    citecolor=blue
    }

\usepackage{placeins}

\usepackage[maxbibnames=99,uniquename=false,uniquelist=false,style=authoryear-comp,giveninits=true,dashed=false,natbib]{biblatex}
\DeclareFieldFormat{pages}{#1} 
\renewbibmacro{in:}{} 
\AtEveryBibitem{\clearfield{month}}

\DeclareFieldFormat{journaltitle}{\mkbibemph{#1}\isdot}
\renewbibmacro*{journal+issuetitle}{%
  \usebibmacro{journal}%
  \setunit*{\addcomma\space}%
  \iffieldundef{series}
    {}
    {\newunit
     \printfield{series}%
     \setunit{\addspace}}%
  \usebibmacro{volume+number+eid}%
  \setunit{\addspace}%
  \usebibmacro{issue+date}%
  \setunit{\addcolon\space}%
  \usebibmacro{issue}%
  \newunit}

\AtEveryBibitem{\clearlist{language}}

\newcommand{\splitatcommas}[1]{%
  \begingroup
  \begingroup\lccode`~=`, \lowercase{\endgroup
    \edef~{\mathchar\the\mathcode`, \penalty0 \noexpand\hspace{0pt plus 1em}}%
  }\mathcode`,="8000 #1%
  \endgroup
}

\usepackage{xr}
\makeatletter

\newcommand*{\addFileDependency}[1]{
\typeout{(#1)}
\@addtofilelist{#1}
\IfFileExists{#1}{}{\typeout{No file #1.}}
}\makeatother

\newcommand{\E}{\mathbb{E}}
\newcommand{\Var}{\mathrm{Var}}
\newcommand{\p}{\mathbb{P}}
\newcommand{\q}{\mathbb{Q}}
\newcommand{\bp}{\textrm{BP}}
\newcommand{\pk}{\textrm{PK}}
\newcommand{\bvrp}{\textrm{BVRP}}
\newcommand{\rv}{\textrm{RV}}
\newcommand{\f}{\mathcal{F}}

\begin{document}

\title{Risk Premia in the Bitcoin Market }

 \author{Caio Almeida\thanks{E-mail: calmeida@princeton.edu. Department of Economics, Princeton University, USA} \and Maria Grith\thanks{E-mail: grith@ese.eur.nl. Erasmus School of Economics, Erasmus University Rotterdam, The Netherlands} \and Ratmir Miftachov\thanks{E-mail: ratmir.miftachov@hu-berlin.de. School of Business and Economics; Institute of Mathematics, Humboldt-Universität zu Berlin, Germany} \and Zijin Wang\thanks{E-mail: wangzijin516@smail.swufe.edu.cn. School of Mathematics, Southwestern University of Finance and Economics, China}}


\date{}
\maketitle
\vspace{-0.5cm}
\begin{abstract}
\noindent 





We analyze the first and second moment risk premia in the Bitcoin market based on options and realized returns and contrast them to the premia embedded in the main US stock index market. First, Bitcoin is much more volatile and has a higher variance risk premium than the S\&P 500. By decomposing the return premium into different regions of the return state space, we find that while most of the S\&P 500 equity premium comes from mildly negative returns, the corresponding negative Bitcoin returns (between three and one standard deviations)  account for only one-third of the total Bitcoin premium (BP). Further, applying a novel clustering algorithm to a collection of estimated Bitcoin option-implied risk-neutral densities, we find that risk premia vary over time as a function of two distinct market volatility regimes. The low-volatility regime implies a relatively high share of BP attributable to positive returns and a high Bitcoin Variance Risk Premium (BVRP). In high-volatility states, the BP attributable to positive and negative returns is more balanced, and the BVRP is lower. These results suggest Bitcoin investors are more concerned about variance and upside risk in a low-volatility regime.



\vspace{0.1in}


\bigskip
\end{abstract}
\newpage

\begin{refsection}

\section{Introduction} \label{sec:introduction}

The cryptocurrency market plays a central role in the digital economy, comprising thousands of cryptocurrencies, numerous exchanges, and a global market capitalization of billions of dollars. As this market expands, financial derivatives linked to cryptocurrencies and crypto-traded funds are gaining popularity. Similar to traditional assets, crypto derivatives can provide essential insight into risk premia, reflecting investors' compensation for taking on risk. Although there is extensive literature on risk premia for traditional assets, particularly equities, there is limited analysis regarding these premia in the cryptocurrency market. This paper aims to fill this gap by conducting a comprehensive analysis of the Bitcoin risk premia using options data, with a focus on the Bitcoin return premium (BP) and the variance risk premium (BVRP).

We start by compiling a reliable joint dataset of Bitcoin returns and option prices to document several key stylized features of the unconditional Bitcoin risk premia. Then, we propose a novel two-stage approach to identify the variations in Bitcoin risk premia and analyze their response to market conditions. In the first stage, we construct a series of risk-neutral density (RND) functions implied by option prices that, jointly with the returns series, enable us to assess the properties of risk premia. In the second stage, we utilize functional data analysis classification tools to classify RNDs into homogeneous clusters that characterize different market regimes.
For these regimes, we calculate the corresponding conditional risk premia (BP and BVRP).

Our analysis focuses on Bitcoin -- the first decentralized and most widely adopted cryptocurrency. 
We view Bitcoin as a digital asset and utilize traditional asset pricing tools to derive insights into its behavior.\footnote {In the U.S., Bitcoin falls under the Commodity Futures Trading Commission (CFTC). Some studies compare Bitcoin with commodities (\cite{Bianchi2020-ci, hou2020pricing, Alexander2021-ai}), while others treat it as a currency (see \cite{Schilling2019-gm, Cao2021-xt, Uhlig2024-qp}).} The fundamental value of Bitcoin is related to net transactional benefits on the platform (\cite{Biais2023-or}), and its price fluctuates substantially due to uncertainty about the fundamentals, institutional risk, sentiment, speculation, or manipulation.\footnote{Cryptocurrencies -- such as Bitcoin -- facilitate peer-to-peer transactions on a digital platform supported by blockchain technology, which uses cryptography—hence the term "crypto." \citet{Biais2023-or} relate Bitcoin price to blockchain technology and its ecosystem. Further studies relate the value of cryptocurrencies to their adoption and evaluate the advantages of tokenization (\cite{Athey2016-ji}, \cite{Cong2021-ma}, \cite{Hinzen2022-jb}, \cite{Sockin2023-tz}, \cite{Sockin2023-wx}, \cite{Hautsch2024-ma}). For studies on price and volume manipulation, we refer to \citet{Makarov2020-xl}, \citet{Griffin2020-fn}, and \citet{Cong2023-fn}; for sentiment and policy uncertainty, refer to \citet{Shen2019-gm} and \citet{Demir2018-kx}.} Many previous studies have explored the factors that influence Bitcoin prices to assess risk premia (\cite{Liu2021-yw}). In this analysis, however, we specifically focus on the risk premia implied by Bitcoin returns and option prices from the Deribit trading platform. Deribit is the largest exchange for crypto-derivatives globally and serves as the primary data source for most existing studies on Bitcoin options (e.g., \cite{Alexander2021-ai}, \cite{Foley2022-py}, \cite{Winkel2023-kk}, \cite{Alexander2023-wq}). 
During our analysis period, Bitcoin's price largely fluctuated independently of traditional macro-financial markets, enabling us to examine this market in isolation.\footnote{\citet{Bianchi2020-ci} reports limited correlations between cryptocurrencies and traditional asset classes, with macro indicators having minimal impact on crypto markets. \citet{Liu2021-yw} finds no significant link between Bitcoin returns and consumption or production growth. \citet{Alexander2021-ai} note that, before the COVID-19 pandemic, Bitcoin's VRP did not align with that of other assets; however, it became highly correlated with the VRP of equity and gold during the pandemic. \citet{Chen2021-ue} show that Bitcoin’s pricing kernel is decoupled from the consumption kernel, with minimal impact from real-economy shocks, like the Covid-19 crisis.} 




Employing semi- and nonparametric statistical techniques on options data and return time series enables us to identify patterns in the Bitcoin market with minimal restrictions and compare them with those of traditional financial assets. We start by calculating unconditional measures of risk premia. 
For a one-month horizon, we find that the BP averages 66\% per annum, while the BVRP averages 14\%, both of which are significantly higher than the premia of traditional assets.\footnote{The point estimate for the first moment premium is known to be a noisy measure. In the paper, we also estimate lower bounds for the BP based on measures extracted from the options data in the spirit of \citet{Martin2017-sl} and \citet{Chabi-Yo2020-tv}. We find that the average lower bound for annualized monthly BP is 66\% for the former and 84\% for the latter.}  For comparison, the average lower bound of S\&P 500 equity premium (EP) for a one-month horizon is 5\% (\cite{Martin2017-sl}). Additionally, focusing on the second moment risk premium, \citet{Bollerslev2009-sb} find a 2\% variance risk premium (VRP) for the S\&P 500. Although Bitcoin demonstrates significantly greater volatility compared to traditional assets like the S\&P 500—by a considerable margin—its annualized Sharpe Ratio (SR) is approximately 0.8, which is nearly twice that of the S\&P 500.\footnote{The SR for the S\&P 500 is approximately 0.5 (\cite{Dew-Becker2021hedging}), and the SR for the SPY is 0.45 per annum (\cite{feng2022factor}). Sections \ref{sec:assets} and \ref{sec:premia} in the Appendix provide a more detailed comparison between Bitcoin and traditional markets.}

Although the unconditional BP can be estimated using Bitcoin returns and interest rates alone, understanding how it varies across different return states necessitates a more comprehensive dataset, which option prices can provide. Using the equity premium (EP) decomposition method proposed by \citet{Beason2022-ht}, we find that large positive returns (ranging from 20\% to 60\%\footnote{Bitcoin returns from 20\% to 60\% correspond to approximately one to three standard deviations of the return distribution.}) account for 39\% of the Bitcoin premium. In contrast, less than one-third of the premium for the S\&P 500 is attributable to positive returns. We find that while most of the S\&P 500 equity premium comes from mildly negative returns, the corresponding negative Bitcoin returns between three and one standard deviations account for only one-third of the total Bitcoin premium (BP).  
These findings suggest that the right side of the Bitcoin return distribution contains crucial information about the market’s key features.\footnote{Focusing on high-frequency data, \citet{scaillet2020high} demonstrate that most price jumps in the Bitcoin market are positive, challenging the common belief that jumps typically indicate impending price crashes. In a separate analysis of options, \citet{Alexander2023-wq} reveal that Bitcoin prices are susceptible to both upward and downward jumps, which in turn affect the shape of the implied volatility curve.}


A natural follow-up question is whether risk premia change with market conditions. To explore this, we use information derived solely from option prices to identify the main Bitcoin market regimes. The data from single-day Bitcoin prices can be unreliable due to a high signal-to-noise ratio. 
In contrast, Bitcoin options are available for various strikes and maturities each day, supplying more reliable information about market pricing. Additionally, their forward-looking nature makes options more sensitive to expected market conditions, making them ideal for capturing signals regarding changes in the market. We propose a novel clustering algorithm for a collection of RNDs estimated from option prices. The objective is to classify them into clusters, allowing the mean density within a cluster to reflect stylized information for asset pricing in a specific market regime. We identify two main clusters: a high-volatility (HV) regime and a low-volatility (LV) regime. In the high-volatility regime, the contribution of large positive and negative returns to the BP is roughly the same, around 30\%, for the selected return segments. 
On the other hand, under the low-volatility regime,  around 52\% of the BP is attributable to large positive returns between 20\% and 60\%. The BVRP is higher in this regime. These results suggest Bitcoin investors are more concerned about variance and upside risk in the calmer market. Our analysis, based on clusters, also unveils a negative relationship between BP and BVRP.

This paper is structured as follows. We start with a brief literature review in the next subsection. Section \ref{sec:data} provides a detailed overview of the data. Sections \ref{sec:methodology} and \ref{sec:estimation} introduce the theoretical framework and the estimation methodology, followed by Section \ref{sec:results}, which highlights the main findings. Section \ref{sec:conclusion} provides a thorough summary of the findings and implications of this article, and offers recommendations for future research.

\subsection{Related Literature}



In this article, we empirically document stylized features of risk premia in the Bitcoin market. We estimate unconditional risk premia implied by options data and returns (BP and BVRP), decompose BP on return states, propose a statistically-based clustering method and track how risk premia vary within these clusters. Our work connects with three main strands of literature.

The first strand analyzes risk premia in cryptocurrencies using Bitcoin options and returns. While studies on cryptocurrency indexes are at a more developed stage, research about derivatives of digital assets is still in its early stages, with limited exploration of cryptocurrency options-implied risk premia.  We contribute to this literature by documenting the stylized features and variation of BP and BVRP across return states and market regimes using an extended dataset. Our unconditional risk premia estimates align with those reported in the existing literature on Bitcoin; i.e., the return premium in \citet{Foley2022-py} and \citet{Wilson2024-cz}, and the variance premium in \citet{Alexander2021-ai}\footnote{\citet{Alexander2021-ai} are the first to construct a Bitcoin VIX Index and estimate VRP for Bitcoin, using options traded on Deribit exchange.} and \citet{Winkel2023-kk}. Relying on our clustering results, we observe a negative relationship on average between BP and BVRP, contrasting with the positive correlation between future returns and VRP reported in the equity market (e.g., \cite{Bollerslev2009-sb}). Focusing on the time-series properties of Bitcoin variance risk premium, \citet{Alexander2021-ai} find that it spikes before large positive or negative returns on Bitcoin.  The atypical behavior of Bitcoin is documented by \citet{hou2020pricing} as an "inverse leverage effect", resembling commodities, which suggests that positive changes in the prices are associated with higher volatility. In a similar vein, but employing a different approach, we notice a positive relationship between returns and volatility.  
\citet{Alexander2023-aw} document an upward-sloping implied volatility curve, particularly during bullish phases of Bitcoin's price. 
We find that the positive right slope of Bitcoin options implied volatility is steeper during the low-volatility regime when the absolute BP is lower, but the BP attributable to positive returns is higher.

The second strand of literature focuses on the decomposition of risk premia on return states. The risk price is naturally expressed as a function of returns through the pricing kernel (PK), and numerous studies have characterized it using options and returns data (e.g., \cite{Jackwerth2000-ed, Ait-Sahalia2000-gs}), with a primary focus on the equity market. We will discuss this literature in detail in what follows. In contrast, the literature on the decomposition of EP on return states is relatively sparse, focusing primarily on equity premia. We contribute to this literature by documenting the first moment premium decomposition and PK patterns for the Bitcoin market.  Our pricing kernel unconditional estimate for a one-month horizon is hump-shaped, featuring an increasing segment for small and medium negative returns of the Bitcoin index. For the Bitcoin market, \citet{Winkel2023-kk} explore the PK term structure and find that empirical PKs are W-shaped for longer-dated options.  Our paper employs the EP decomposition of \citet{Beason2022-ht} using refined non-parametric techniques and documents the differences between the S\&P 500 and Bitcoin markets. Relying on the same methodology, \citet{Almeida2024-dt} find evidence for the prominent contribution of right-tail return states to the EP and VRP of the S\&P 500 Index for options with ultra-short maturities. A complementary approach for studying risk premia for different return segments is proposed by \citet{chabi-yo2023decomposition}, who find that EP and VRP of S\&P 500 Index are largely driven by the left tail of the return distribution. Our results show that most of the BP is attributable to positive returns, while the negative returns contribute mostly to the BVRP.


The third strand of literature relates to the conditional estimation of options-implied risk measures. Identifying the relevant factors for conditioning is particularly important in emerging markets like Bitcoin, which lack a clear understanding of the risks that contribute to the overall risk premia.  Previous research on traditional assets has focused on how observable factors — such as the business cycle, volatility index, and news — affect various risk measures. By contrast, our paper introduces a data-driven approach to identify the drivers and estimate conditional premia that is both economically and statistically sound. We specifically investigate the information found in option prices to reveal relevant information for conditioning. We estimate nonparametrically the risk-neutral density (RND) implied by options in liquid markets (\cite{Breeden1978prices}).\footnote{In contrast, estimating the conditional physical density without making restrictive assumptions is widely recognized as challenging (\cite{Shumway2018pricing}, \cite{Barone-Adesi2020-ck}).}
Our novel contribution to the existing literature lies in applying established clustering statistical techniques for functional data (\cite{Jacques2014-hx}) within the context of financial data analysis. We find that variance (both BRV and BVIX) is an important variable for characterizing the drivers of the clusters. This finding aligns with several other studies that consider volatility as an important conditioning variable.
For instance, \citet{Chabi-Yo2012-kh} demonstrate that PK increases with market volatility, while \citet{Song2016-pr} and \citet{Shumway2018pricing} show that the PK remains nonincreasing when consistently conditioned. 
Our work is also related to the conditional estimation of the pricing kernel and its link to the volatility risk premium. In the equity index markets, \citet{Christoffersen2013-dw} propose a U-shaped pricing kernel driven by a positive VRP, while \citet{Almeida2022-gd} use demand as a state variable and find that the pricing kernel is U-shaped and VRP is high when public investors are net-selling OTM options. \citet{Grith2017-dj} report a hump-shaped pricing kernel under low VRP and a U-shaped pricing kernel under high VRP. In contrast, our findings in the Bitcoin market reveal a more pronounced hump in the LV regime, which is characterized by a higher BVRP. Furthermore, our research suggests that factors beyond volatility may play a significant role in explaining the variation in risk premia within the Bitcoin market.\footnote{For instance, focusing on equity indexes, \citet{Rosenberg2002-ox} find a positive correlation between the PK's slope and recession indicators, while it negatively correlates with expansion signs for S\&P 500 Index, while \citet{Grith2013-uu} observe that the PK hump is more pronounced during economic expansions than in recessions for German DAX Index. A similar analysis focusing on fluctuations in the cryptocurrency market can also be performed for Bitcoin.} 


\section{Data} \label{sec:data}

The data contains cash-settled European-style options traded on the Deribit exchange and daily Bitcoin prices, available via the \href{https://blockchain-research-center.com/}{Blockchain Research Center} (BRC). Deribit is a margin-trading platform specializing in Bitcoin and Ethereum options, as well as futures and options. It has emerged as the leading cryptocurrency options exchange globally. This platform was established in the Netherlands in June 2016 and was registered in Panama and then Dubai. In May 2025, it was acquired by Coinbase. 
 Daily USD-denominated Bitcoin prices are collected after the early-adoption phase from January 2014 to December 2022 from Deribit.\footnote{We discard data prior to 2014 when Bitcoin prices were very volatile and less reliable. For instance, Bitcoin skyrocketed from \$13 at the beginning of 2013 to \$1000 by November of this year, increasing by over 75 times in just 11 months.} Deribit calculates Bitcoin settlement price as a weighted average across eleven major cryptocurrency exchanges over the last 30 minutes before settlement time (8 am UTC). Bitcoins are divisible, such that the quantity traded can be expressed in decimals, and are traded around the clock on several exchanges. The trading fees on Deribit are 0.03\% of the underlying or 0.0003 Bitcoin per option contract, capped at 12.5\% of the contract's value. 

We use daily transaction-level options data spanning from July 2017 to December 2022. The raw data includes the timestamp, order type (call or put), volume, instrument price, strike price, spot price (the price of the underlying), implied volatility, and transaction type (buy or sell). Each contract has a lot size of 1 $\mathrm{Bitcoin}$. All prices and instruments are denominated in U.S. Dollars. 
We implement some filtering to mitigate potential errors in the raw data. A notable distinction in our paper is the nature of transaction data, which provides a single option price per transaction rather than separate bid and ask prices. Further, we exclude option observations where i) options whose implied volatility is zero or negative, ii) implied volatility is missing or non-positive, iii) no-arbitrage conditions are violated, 
iv) transaction quantity is non-positive.  
After filtering, our dataset comprises 1992 days, including 5,384,537 transactions. 

In our analysis, we focus on put and call options that have 27 days to expiration, which we refer to as monthly options. However, on Deribit, options with this specific maturity are not always available every day. To address this, we employ an interpolation scheme using the nearest maturities. For clustering purposes, we consider a more comprehensive data set that contains information about the term structure of implied volatility curves. Specifically, we additionally require smooth implied volatility curves with 9, 27, and 45 days to expiration. For this, we effectively utilize options with expiration dates ranging from 3 to 120 days. 
These options always expire on a Friday.
We define time-to-maturity $\tau$ measured in years for each option contract. The moneyness of a contract is $m = K/S$, where $K$ is the strike price, and $S$ denotes the current Bitcoin price. We use the daily 1-month Treasury bill rate available on the FRED (Federal Reserve Bank of St. Louis) website as a proxy for the risk-free rate, and we set the cost-of-carry rate to zero.\footnote{We have also experimented with the zero interest rate --  in accordance with the practices observed on the Deribit exchange -- as a proxy for the risk-free rate. This choice makes virtually no difference to our calculations.}

\begin{center}
\begin{threeparttable}
\centering \footnotesize
\caption{\footnotesize  Summary statistics of Bitcoin options}\label{tab:summary_option}
\begin{tabular}{L{0.1\textwidth} C{0.1\textwidth}C{0.1\textwidth}C{0.1\textwidth}C{0.1\textwidth}C{0.1\textwidth}C{0.1\textwidth}C{0.1\textwidth}C{0.1\textwidth}C{0.1\textwidth}C{0.1\textwidth}}
\toprule
        & \multicolumn{3}{c}{Call} & \multicolumn{3}{c}{Put}\\
\cmidrule(r){2-4}\cmidrule(r){5-7}
  & TTM & Moneyness & IV & TTM & Moneyness & IV\\
\cmidrule(r){2-7}
Mean      &  29.27 &  1.21 & 0.82 &  24.57 &  0.91 & 0.89 \\ 
Median    &   9    &  1.06 & 0.77 &   8    &  0.94 & 0.82 \\ 
Std. Dev. &  49.95 &  0.55 & 0.29 &  44.08 &  0.19 & 0.37 \\ 
Min       &   1    &  0.08 & 0.05 &   1    &  0.07 & 0.1  \\ 
Max       & 372    & 17.71 & 5    & 372    & 15.67 & 5    \\ 
\bottomrule
\end{tabular}
\renewcommand{\baselinestretch}{0.8}\footnotesize
\begin{minipage}{0.88\textwidth}
The table gives a summary statistic of the filtered Bitcoin call and put options traded daily from July 1, 2017, to December 17, 2022. It showcases the option characteristics, such as the time to maturity (TTM), moneyness, and the implied volatility (IV) from Deribit. The number of transactions for call options amounts to 3,940,541 and 3,468,020 for put options. Consequently, our dataset comprises 1,301 days that include a total of 7,832,590 Bitcoin option transactions, with a daily average transaction volume of 3,721 option contracts. 
\end{minipage}
\end{threeparttable}
\end{center}

The summary statistic presented in Table \ref{tab:summary_option} highlights the essential option characteristics, including time-to-maturity (TTM), moneyness, and implied volatility (IV) from Deribit. 
Similar to \citet{teng2022financial}, we find that the range of moneyness in the Bitcoin options market is significantly wider than that of traditional options markets, which can be attributed to the highly volatile nature of Bitcoin.\footnote{\citet{Liu2021-yw} find that Bitcoin returns exhibit significantly higher volatility compared to stocks, ranging from 5 to 10 times greater depending on the investment horizon.} The average Bitcoin IV level of 82\% is much higher than the average S\&P 500 IV level.
Furthermore, options with shorter tenors are more frequently traded than those with longer tenors. 
Before 2020, the average daily transaction volume was approximately 646 contracts; after 2020, this number increased to 3,721 contracts, which is almost a 476\% increase.\footnote{A similar trend for average daily transactions is observed in SPX options, albeit the increase is only around 20 \%, from 921,948 contracts before 2020 to 1,109,514 contracts after 2020. We display the average daily Bitcoin options in Figure \ref{fig:transaction_daily} in the Appendix.} More information about the data can be found in Appendix \ref{app:data_analysis}. 

\section{Theoretical Framework} \label{sec:methodology}

Let the price process of the Bitcoin Index be a stochastic process with continuously distributed marginals $S_t$ under the physical measure $\p$, equipped with a filtration $\f_t$. In what follows, we focus on unconditional distribution of the $\tau$-days ahead random returns $R = (S_{t+\tau}-S_t)/S_t$.\footnote{Working with simple net returns enables a direct comparison to \citet{Beason2022-ht}.} Removing the time index indicates that we are not adopting a time-series approach.

The arbitrage-free assumption implies the existence of an equivalent measure $\q$ (to $\p$) identified with a risk-neutral pricing rule. Under such a measure, discounted prices have the martingale property, such that the returns satisfy $\E_{\q}(R)=R^f$, with $R^f$ being the risk-free rate. Furthermore, we assume that the probability measures $\p$ and $\q$ are differentiable with respect to the returns. Then, for each value $r$ of the returns $p(r)=\frac{\partial \p(r)}{\partial r}$ and $q(r)=\frac{\partial \q(r)}{\partial r}$, with $q(r)$ being the risk-neutral density and $p(r)$ being the physical density. In addition, we assume that $p(r) \neq 0$ for all possible returns $r \geq -1 $. We utilize the information contained in the two pricing rules to analyze the first  and second moment risk premia and gain insights into the pricing of risk.

Under no-arbitrage, it holds that the unconditional return premia for Bitcoin (BP) is
\begin{equation}\label{eq:ep}
    \bp  \coloneqq \mu_{\p} - \mu_{\q}, 
\end{equation}
where $\mu_{\p}=\E_{\p}(R)$ and $\mu_{\q}=\E_{\q}(R)$. A positive BP premium is typically associated with investors seeking compensation for assuming directional risk. Equation \eqref{eq:ep} allows for a direct comparison with \citet{Beason2022-ht}, with one significant difference: Bitcoin does not provide dividends. 

To characterize risk pricing consistently with a more complex data-generating process that accommodates stochastic variance and jumps, we analyze the variance risk premium for Bitcoin (BVRP)

\begin{equation}\label{eq:VRP}
\bvrp \coloneqq \sigma^2_{\q} - \sigma^2_{\p},
\end{equation}
where $\sigma^2_{\q}= \Var_\q(R)$ and $\sigma^2_{\p}= \Var_\p(R)$. 
A positive BVRP indicates that variance buyers are willing to pay a premium to hedge away upward movements in the variance of the returns. In contrast, a negative BVRP indicates that the buyers would request a positive premium to long volatility. The BVRP in Equation \eqref{eq:VRP} is defined differently than the BP in that a premium is paid to avoid the variance risk of the asset, hence the resulting sign inversion for the moments under the two measures. 


\subsection{Bitcoin Premium Decomposition}

We utilize the method proposed by \citet{Beason2022-ht}, originally used to analyze the S\&P 500 market, to investigate the decomposition of BP in different return states, such that 
\begin{equation}\label{eq:epx}
\bp(r) = \frac{\int_{-1}^r x \{p(x)-q(x)\}dx}{\bp}
\end{equation}
measures the fraction of the average Bitcoin first moment premium that is associated with returns below $r$. 
For ease of interpretation, we use a standardization by the BP 
that guarantees that the $\bp(r)$ function approaches zero for returns in the far left tail and one for returns in the far right tail. Note that the $\bp(r)$ function is not restricted to be monotonically increasing and can take values larger than one. Equation (\ref{eq:epx}) indicates that $\bp(r)$ increases when the risk-neutral density exceeds the physical density for negative return states, and when the physical density is greater than the risk-neutral density for positive return states. Economically, the increasing segments of $\bp(r)$ are associated with states that contribute positively to the return premium.

The no-arbitrage assumption is also equivalent to the existence of a positive random variable $\pi$, called a stochastic discount factor (SDF), such that $\E_{\p}(R \pi)=R^f$. We refer to the projection of the SDF $\pi$ on the set of Bitcoin returns as the pricing kernel function $\pk\left(r\right) =\E\left[\pi |R=r\right]$ with $\E[\pk\left(R\right)]=1$. We define the pricing kernel associated with return $r$ as the Radon-Nykodim derivative of the risk-neutral measure with respect to the physical measure
\begin{equation}\label{eq:pk}
    \pk(r)=\frac{\partial \q}{ \partial \p}(r) =  \frac{q(r)}{p(r)}. 
\end{equation}
\pk(r) reflects the risk price associated with the return $r$. Whenever the risk-neutral density exceeds the physical density, PK is larger than one, which means that the market risk price for that state is expensive relative to its (state) probability.  Consequently, PK can provide crucial insights into the pricing rules of risk for the marginal investor across different return segments.\footnote{Most studies define the SDF in general equilibrium models as a function of aggregate consumption or wealth, often assuming a representative agent with expected utility preferences. A common assumption is that SDF is proportional to the marginal utility, decreasing as consumption rises. Consequently, under certain conditions, such as co-monotonicity between consumption and returns, the pricing kernel $\pk(r)$ is expected to decrease with returns.}


The shapes of the $\bp(r)$ and $\pk(r)$ functionals are closely related and provide insight into the entire distributions of returns. Both functions can provide direct information about the BVRP, risk price, and risk aversion. A connection between the $\pk(r)$, the shape of the $\bp(r)$, and the prices of risk for monotonically decreasing PK focusing on the negative returns is provided by \citet{Beason2022-ht}. Further, \citet{Almeida2024-dt} document the link between the $\pk(r)$ and $\bp(r)$ with a particular focus on positive returns for both monotonic and non-monotonic pricing kernels. \citet{Schreindorfer2025-fa} analyze risk aversion and pricing kernel focusing on negative market returns and their relation to the variance of returns. 

\subsection{Market Regimes and Risk-Neutral Density Clustering}
\label{sec:cluster}
In this article, we propose a nonparametric, data-driven approach for the conditional analysis of Bitcoin premia and the pricing kernel shape, highlighting their variations across different market regimes. We utilize risk-neutral densities to capture future market expectations and to identify similar market regimes.  Our objective is to group these densities into homogeneous clusters. Due to the continuous nature of the data, we adopt a functional data approach and integrate information from both moneyness and time-to-maturity dimensions relevant to option prices.\footnote{Functional data consists of continuously observed curves. For a general overview of functional data, we refer to \citet{Wang2016-zr}. For more information on multivariate functional data methods, see \citet{Koner2023-cz}.} 

A straightforward method for clustering densities involves focusing on a specific time to maturity, which we refer to as the \textit{univariate} clustering approach. While this method is simple to implement, it has one significant drawback: it overlooks the information contained in the term structure of options. This limitation can be addressed by using a more reliable method known as the \textit{multivariate} clustering approach, which produces economically meaningful results. The \textit{multivariate} approach considers a set of time-to-maturities (either continuous or discrete) that represent the term structure and treats the corresponding risk-neutral densities as a series of curves. This method is preferred over the \textit{univariate} approach because it incorporates information across both the moneyness and expiry dimensions, leading to more reliable clustering outcomes. Additionally, it allows for a more nuanced representation of investors' future expectations.

Although the implied volatility surface has the potential to be used for clustering, our empirical investigation indicates that employing risk-neutral densities (RNDs) leads to more stable clustering outcomes.\footnote{Clustering results based on implied volatilities are available upon request.} This stability can be attributed to the fact that RNDs are closely connected to the second derivative of the call function for a fixed maturity (as discussed in Section \ref{sec:rnd}). Prior research in functional data analysis shows that second derivatives are more responsive to changes in the latent processes that drive variability in the observed data (\cite{Grith2018-qr}). In contrast, implied volatilities involve a simpler nonlinear transformation from option prices to volatilities. 

To effectively differentiate between these functional objects, one needs to compute a distance metric and then apply a clustering algorithm.  In the context of clustering and classifying functional data, hierarchical clustering and the $k$-means partitioning methods are two classical and widely used techniques; see \citet{Jacques2014-hx} for a review on clustering. We use the default \(L^2\) distance in the multivariate function space to apply hierarchical clustering, where the mean functions are defined as the centers of each cluster. 

However, because density functions satisfy the constraints $\int f(x) d x=1$ and $f \geq 0$, they are not situated within a vector space. Consequently, traditional functional data analysis methods based on Hilbert space are not applicable \citep{petersen2016functional}. An isomorphic mapping of the densities to the standard $L^2$ space is required to perform standard statistical Hilbert space methods. Different transformations are possible, such as taking the natural logarithm. As outlined in \citet{machalova2016preprocessing} and \citet{eckardt2022generalised}, a straightforward isomorphism that has shown better results in practice is the centered-log-ratio (CLR) transformation. The transformation is applied to the RND function and is defined as 
\begin{align}
     \text{clr}\{q(r)\} = \log\left\{\frac{q(r)}{\mu_{G}}\right\}, \label{eq:gm}
 \end{align}
with the geometric mean of the risk-neutral density function given by $\mu_{G} = \exp\left[  \E\{  \log(q(r) ) \} \right]$.

In the second step, we compute the $L^2$ distance between all pairs of transformed density sets indexed by $i$ and $j$, for a continuum of $\tau$. The distance between two-dimensional functions is defined as:
 \begin{align*}
        D(i,j) = \sqrt{\int_{\tau}\int_r\left[\text{clr}\left\{q_{i}(r,\tau)\right\} - \text{clr}\left\{q_{j}(r,\tau)\right\} \right]^2 dr d\tau}\ \ \ \ \text{for all} \ i,j, 
    \end{align*}
where $i=1,\ldots,T$ and $j=1,\ldots,T$.
Building upon the resulting Euclidean distance matrix, the risk-neutral densities are grouped into homogeneous clusters, where homogeneity is measured by the symmetric distance measure of the transformed densities. 

 We apply the agglomerative hierarchical clustering method with the Ward linkage on the calculated Euclidean distance matrix \citep{ward1963hierarchical}. The Ward method, which minimizes the overall within-cluster variance, has the advantage of producing well-balanced clusters. Moreover, the obtained clusters are also robust with respect to the choice of linkage (complete, single, or average).\footnote{More details on the agglomerative clustering method can be found in Chapter 14 of \citet{hastie2009elements}.} 

\section{Estimation Procedure}\label{sec:estimation}

To unify the various methods in existing research that combine option prices, underlying asset prices, and investment decision data, we employ flexible estimation procedures. All our estimators are semiparametric or nonparametric. This approach enables us to better understand the underlying phenomena without imposing rigid models.

\subsection{Risk Premia Estimation}\label{sec:estpremia}

\noindent\textbf{Empirical Bitcoin Premium: $\widehat{\bp} = \widehat{\mu}_{\p} - \widehat{\mu}_{\q}$}.  It is the unconditional premium involved in estimating the first moment of the returns. The point estimate for the first moment premium is known to be a noisy measure. Therefore, we have experimented with several estimators. For the first annualized empirical moments under the $\p$ measure, the following estimator gives more stable results $\widehat{\mu}_{\p}=\frac{365}{27} \int_{-1}^{1}x\{\hat{p}(x)\}dx$ and reported in what follows. Several other approaches are possible. Results for the sample mean are available in Table \ref{tab:SR}. We also computed the option-based first moment lower bounds following \citet{Martin2017-sl} and \citet{Chabi-Yo2020-tv}. As discussed in the data section, we use the 1-month (annualized) Treasury bill rate as a proxy for the risk-free rate.

\textbf{Empirical Variance Risk Premium: $\widehat{\operatorname{VRP}} =\widehat{\sigma}^2_\q -\widehat{\sigma}^2_\p $}. Estimating the variance risk premium is a primary focus of this work. Building on earlier studies, this work utilizes option-based volatility as a proxy for the $\q$-volatility. We calculate the daily volatility Index, the BVIX, for different tenors. Our methodology for constructing the daily BVIX$_t$ utilizes intraday option data on Bitcoin. It builds on the fair pricing of variance swaps employed by the CBOE to compute the Volatility Index (VIX). The squared BVIX reflects a market-specific expected Bitcoin variance directly captured from options, and we use its sample mean $\text{BVIX} = \frac{1}{T}\sum_{t=1}^T \text{BVIX}_t$ as a proxy for the $\q$-variance. Details on the calculation of the BVIX are provided in Appendix \ref{app:BVIX}. To estimate the unconditional physical centered second moment, we first obtained a series of (annualized) monthly realized variances 
$\text{RV}_t=\frac{365}{27}\sum_{l=1}^{\tau}r_{d,t-l}^2,\,r_{d,t}=\log{S_t/S_{t-1}},$ calculated as the sum of squared log returns over the past $27$ days, and then we take the mean $\text{RV} = \frac{1}{T}\sum_{t=1}^T \text{RV}_t$ over the entire sample.\footnote{As an alternative approach, variance is estimated, by integrating the squared deviation of returns from the mean over the respective risk-neutral density or physical density for a specified maturity $\widehat{\sigma}_\q^2=\frac{365}{\tau}\int_{-1}^\infty \left\{ x - \int_{-1}^\infty z\hat{q}(z)dz \right\}^2\hat{q}(x)dx$ and $\widehat{\sigma}_p^2=\frac{365}{\tau}\int_{-1}^\infty \left\{ x - \int_{-1}^\infty z\hat{p}(z)dz \right\}^2\hat{p}(x)dx.$ All volatility measures are annualized for consistency. These results are available upon request.} 

\noindent\textbf{Bitcoin Premium Decomposition and Pricing Kernel}. The Bitcoin premium decomposition is given by Equation \eqref{eq:epx}. We implement it for the annularized monthly returns. The pricing kernel is estimated following Equation (\ref{eq:pk}) using the average risk-neutral density \textbf{$\hat{q}(r)$} as defined in Equation (\ref{eq:av_q}) and the physical density $\hat{p}(r)$ defined in Equation \eqref{eq:epdf} on the full sample if overlapping monthly returns.



\subsubsection{Risk Neutral Density}
\label{sec:rnd}
\textbf{Interpolation of Implied Volatility Surface}. The estimation of the risk-neutral density (RND) involves several carefully designed steps, outlined below. First, we need to estimate a smooth implied volatility surface. To accomplish this, we employ a smoothing technique tailored to the data structure, which leverages semi-parametric models. These models impose no-arbitrage restrictions and provide reasonable extrapolation to the tails, where observations are sparse. Second, the smooth surfaces are adopted to estimate a daily set of RNDs for the desired option expiry dates.

To estimate smooth surfaces, we aggregate transaction-level data for both put and call options observed daily, using an unbalanced design that varies in terms of strike prices and time to maturity. Each day, the option data is organized as a series of strings, with each string representing a different time to maturity. Options within a string have a unique time to maturity but feature different strike prices (or levels of moneyness). Observed maturities fluctuate daily. 

Following \citet{Almeida2024-dt}, we utilize options-implied volatilities of the same maturity to fit the parametric Stochastic Volatility Inspired (SVI) model proposed by \citet{gatheral2004parsimonious} on a daily basis. The SVI model is relatively flexible in replicating various shapes of the implied volatility curve. The implied variance can be expressed as follows:
\begin{align*}\label{iv_main}
\omega_{\theta}(r) = a + b \left[ \rho \left\{r - m\right\} + \sqrt{\left\{r - m\right\}^2 + \sigma^2} \right],
\end{align*}
where $r$ denotes the simple return (i.e., $K/S_t-1\mid K =S_{t+\tau}$), and $\theta = [\alpha, \beta, \rho, m, \sigma]$  are parameters that capture various characteristics of the volatility smile, such as its level, slope, and curvature. Similarly to \citet{gatheral2004parsimonious}, to enforce no-arbitrage we impose the constraints that $b>0$, $1 - \left| \rho \right| > 0$, $a + b \cdot \sigma \sqrt{1 - \rho^2} > 0$, and $\sigma > 0$. We estimate the parameter vector $\theta$ 
by minimizing the root mean squared error (RMSE) 
\begin{align*}
\hat{\theta}_{t\mid \tau}=\underset{\theta}{\arg \min } \sqrt{\frac{1}{N_{t,\tau}} \sum_{i=1}^{N_{t,\tau}}\left\{\omega_{t, i}-\omega_{\theta}\left(r_{t, i}|\tau\right)\right\}^2},
\end{align*}
where $\omega_{t, i}$ is the squared observed IV at day $t$ corresponding to an option with maturity $\tau$ and strike indexed by $i \in \{1,\cdots, N_{t,\tau}\}$. 



After generating the SVI-smoothed IV curves for each available maturity on a given day, we interpolate linearly along the maturity dimension--using a weighted average of the two closest maturities to construct IV curves of unobserved maturities.\footnote{This approach to smoothing the surfaces is more flexible than the one used by \citet{Beason2022-ht}, who assume linearity in the SVI parameters with respect to $\tau$ and fit the model by pooling the observations along the moneyness and time-to-maturity dimensions. 
Further details about the interpolation are provided in Appendix \ref{app:interpolation_iv}, and the interpolated IV surface is shown in Figure \ref{fig:IV_surface}.}
After the smoothing step, we fix a discrete set of maturities, focusing on the short leg of the options term structure, with 9, 27, and 45 days to maturity, for which we estimate the RNDs. 





\textbf{Estimation of the Risk-Neutral Density}. In the next step, we rely the classic theoretical result by \citet{Breeden1978prices} to recover the RND for a given maturity of options. 
Through a change of variable, the RND is represented as a function of returns  
\begin{equation}
    q_t(r)=\frac{q_t(K)}{\partial r/\partial K}=e^{r_f\tau} S\frac{\partial^2 C_t}{\partial K^2}.\label{eq:BS}
\end{equation}
Nonparametric estimation of the risk-neutral density (RND) requires a smooth call function based on the strike price. In our implementation, we utilize the smooth implied volatility surface obtained in the previous step to convert the implied volatilities into call prices using the Black-Scholes pricing formula. We then calculate the second derivative numerically to derive the RND estimates. Finally, for $\tau$, we collect a series of daily estimated RNDs indexed by $t$, $\hat{q}_t(r)$. Then the unconditional risk-neutral density is defined as a sample average over the entire sample
\begin{align}
    \hat{q}(r)=\frac{1}{T}\sum_{t=1}^T \hat{q}_t(r) \label{eq:av_q}.
\end{align}

\subsubsection{Physical Density}\label{sec:P-density}

The physical density is estimated as a smoothed empirical probability density function (PDF) of returns. As a first step, the empirical PDF is estimated as a histogram of the full sample of overlapping returns.  Following \citet{Beason2022-ht}, we smooth the empirical PDF between the 10th and 90th return percentile using a 10th-order polynomial. For the tail regions, the Generalized Extreme Value (GEV) distribution is employed. 
The tails of the physical densities are estimated using the generalized extreme value (GEV) distribution following \citet{Figlewski2008estimating}. The GEV distribution function is defined by 
\begin{equation*}\label{eq:gev}
    F_{GEV}(x)=\exp \left\{-\left(1+\xi \frac{x-a}{b}\right)^{-\frac{1}{\xi}}\right\}.
\end{equation*}
Specifically, the discrete empirical density is obtained via a histogram, and then a 10th-order polynomial is used to obtain the continuous density for the density between quantiles 10 and 90. The GEV distribution is used to fit the tails of the left and right parts that are below the 10th quantile and above the 90th quantile. For each tail, three parameters need to be estimated. This requires two points to construct a loss function by making the Probability Density Function (PDF) and Cumulative Distribution Function (CDF) values at these two points close to the values of the empirical density. After all, the estimated density must integrate to one. Thus, the final estimator is

\begin{equation}\label{eq:epdf}
\hat{p}(r)=\left\{
\begin{array}{ll}
    \hat{f}^{GEV}_l(r), &r\le r_{0.1},\\
    \hat{f}_{pol|hist}(r), & r_{0.1}<r<r_{0.9},\\
    \hat{f}^{GEV}_r(r), &r\ge r_{0.9},
\end{array}
\right.
\end{equation}
where the smoothed version of the histogram estimator is denoted by $\hat{f}_{pol|hist}$\footnote{Choosing between 8 and 13 histogram bins has a negligible impact on estimate when the number of bins within 8-13, and we use 12 bins.}, $\hat{f}^{GEV}_l(r)$ and $\hat{f}^{GEV}_r(r)$ are the left and right tails estimated by the GEV distribution, respectively. The 10th and 90th percentiles are denoted as $r_{0.1}$ and $r_{0.9}$, respectively.\footnote{Additionally, robustness evaluations using kernel density estimation yield results that are largely consistent with those derived from the empirical PDF, and are available upon request.}  

\subsection{Conditional Estimation of Risk Measures}

The unconditional estimators presented in Section \ref{sec:estpremia} rely on averaging time series quantities. Similarly, for the conditional estimators, we average observations belonging to each cluster. While the conditional estimates at any particular point are expected to be noisy, the unconditional and conditional ones based on clustering are expected to be more stable. Relying on two clusters that describe highly volatile (HV) and less volatile (LV) market regimes, as shown in Section \ref{sec:clusterRND}, we define conditional estimators below.\footnote{The hierarchical clustering algorithm used indicates two main clusters. We display the two main modes of variation and cluster allocation of the densities in a low-dimensional representation in Section \ref{app:umap} in the Appendix.} 

The conditional (on clusters) variances are $\mathrm{BVIX}_i^2 = \frac{1}{\left|C_i\right|} \sum_{t \in C_i} \mathrm{BVIX}_t^2$ and $\text{RV}_i^2 = \frac{1}{|C_i|}\sum_{t \in C_i} \text{RV}_t$, where each cluster $i$ is represented as a set of dates $C_i=\{ t | t\in \text{Cluster } i \}$ for $i=HV, LV$. Similarly, the risk-neutral density for the clusters is given as
$ \hat{q}_i(r)=\frac{1}{|C_i|}\sum_{t\in C_i} \hat{q}_t(r),\, \text{for}\ \ \ i=HV,LV.$ The conditional $\p$ density is estimated using rescaled returns. Specifically, the rescaled returns are obtained from the full sample overlapping returns according to the volatility levels in each cluster and standardized by the unconditional volatility, as denoted by 
$r_{t,i}=\frac{\text{RV},i}{\text{RV}}r_t,\text{ for cluster } i=HV,LV.$
Finally, $\hat{p}_i(r)$ is computed via Equation \eqref{eq:epdf} using rescaled returns $r_{t,i}.$ We use these densities to retrieve conditional estimates $\widehat{\mu}_{\p, i}$, $\widehat{\bp}_i$ and $\widehat{\pk}_i$.


\subsection{Clustering of Risk Neutral Densities}\label{sec:clusterRND}




The proposed classification is based on the endogenous variation of risk-neutral measures.  We first estimate the risk-neutral density as described in Section \ref{sec:rnd} and take the CLR transformation of Equation \eqref{eq:gm}. For the multivariate clustering approach introduced in Section \ref{sec:cluster}, only the dates for $\tau = 9, 27$, and $45$ are selected, at which all three time-to-maturities of interest are observed. Furthermore, the choice of the number of clusters is underscored by visualizing the risk-neutral densities and the distance matrix in a low-dimensional graph. The first two principal components of the distance matrix are illustrated in Figure \ref{pca}.
Second, the Uniform Manifold Approximation and Projection (UMAP) technique \citep{mcinnes2018umap} is applied, which absorbs non-linear dependencies between the risk-neutral densities. It is elaborated in more detail in Appendix \ref{app:umap}. By marking the reduced-form quantities with the respective cluster, the robustness of the clustering results is confirmed. The UMAP results are illustrated in Figure \ref{umap}. As the low-dimensional structure of the risk-neutral densities as well as the distance matrix indicates, selecting two clusters is indeed a reasonable choice.

To enhance the interpretability of the resulting clusters, we run a logistic regression of the cluster labels on the first four moments of the risk-neutral densities at each day. The regression results are included in Table \ref{tab:logistic_1to4moment_ttm27}. As expected, the coefficients of the moments are highly significant. In particular, a higher variance increases the probability of being in the high-volatility cluster. On the contrary, a higher mean, skewness, and kurtosis\footnote{To check for robustness, we estimated Gaussian tails of the risk-neutral density instead of the GEV distribution and reran the logistic regression. It shows that neither clustering results in the section above nor the results of Table \ref{tab:logistic_1to4moment_ttm27} change significantly.} 
are associated with a higher probability of being in the low volatility cluster. It shows that the variance explains most of the variation in the clusters with an $R^2$ measure of $69 \%$, compared to the other moments. Even if we run a multivariate regression on all moments jointly, it barely increases the explained variation in the clusters. This association gives us reason to refer to the first cluster as the \textit{high volatility} (HV) cluster, and the second cluster as the \textit{low volatility} (LV) cluster.

\section{Empirical Results} \label{sec:results}



Our research focuses on a 27-day investment horizon. Table \ref{tab:RP_2cluster} summarizes the main results for the unconditional and conditional BP and BVRP estimates. The BP is significantly higher than that of traditional investment assets such as currencies, commodities, and stocks, averaging around {66}\% per year. It is known that the point estimate for the first moment risk premium is a noisy measure. For this reason, we also estimate first moment premium lower bounds for the BP based on measures extracted from the options data. The lower bounds reflect the minimum compensation investors require for holding Bitcoin.  The BP lower bound of \citet{Martin2017-sl} requires only knowledge of the risk-neutral variance, while the lower bound of \citet{Chabi-Yo2020-tv} employs higher order moments (skewness, kurtosis) as well. We find that the average lower bound for annualized monthly BP is $65.64 \%$ for the former and $84.78 \%$ for the latter. The annualized risk-neutral and physical monthly variances, proxied by the squared Bitcoin Volatility Index (BVIX) and the realized variance (BRV), are significantly high: 0.72 and 0.58, respectively. The corresponding variance risk premium is 0.14, much higher than that of the S\&P 500 Index—approximately 2\%, according to \citet{Bollerslev2009-sb}.

We further analyze conditional estimates across market regimes. Our results indicate that BP is higher in the HV regime. \citet{Martin2017-sl} lower bounds are estimated at 78.16\% in the HV regime and 43.99\% in the LV regime, while \citet{Chabi-Yo2020-tv} bounds are 102.27\% in the HV regime and 54.51\% in the LV regime. While our point estimates of BP in the LV regime appear reasonable, the lower bounds by both methods indicate that in volatile markets, the BP may be higher than our point estimate. Furthermore, risk-neutral and physical variances exhibit substantial variation across clusters. Specifically, the HV cluster describes a highly volatile market, identifiable by high second moments of Bitcoin returns, where the monthly annualized variances are 0.86 for the risk-neutral and 0.74 for the physical measure. In contrast, the LV cluster describes a less volatile market, identifiable by smaller variance proxies, with a risk-neutral variance of 0.46 and physical variance of 0.29. The overall increased volatility leads to higher option price premia on average across all moneyness levels.\footnote{Higher option prices in the HV regime are also shown as a higher implied volatility surface in Figure \ref{fig:IV_surface} in the Appendix.}
The variance under the two probability measures is quite different and introduces a substantial VRP in both market regimes. 
Surprisingly, the low volatility cluster is characterized by a higher VRP of 0.17 compared to the high volatility cluster of 0.12, suggesting a potential disconnect between variance and VRP.\footnote{\citet{Ait-Sahalia2024-jv} find a disconnect between uncertainty and volatility, and show that equity premium increases with mounting uncertainty. In contrast to their paper, we analyze the variance risk premium for high and low volatility clusters and do not consider a regression framework.} By relying on the average values of the premia within each cluster, we observe a negative relationship between the  BP and BVRP in the Bitcoin market. This contrasts with the findings in S\&P 500 Index market, where a positive relationship between variance risk premium and future returns has been reported by \citet{Bollerslev2009-sb} in a regression setup. 
Simultaneously, our clustering results indicate a positive relationship between returns and variance, supporting the inverse leverage effect in the Bitcoin market documented by \citet{hou2020pricing}.\footnote{\citet{hou2020pricing} find a positive correlation coefficient between the increments in the return and volatility equations of the stochastic volatility with a correlated jump (SVCJ) model of \citet{Duffie2000-tg}. This implies that increasing prices are associated with increasing volatility. We show that the first moment lower bounds in the Bitcoin market, which indicate the minimum compensation for holding Bitcoin, rise following periods of high realized variance. This suggests that BP increases when volatility rises.}

\begin{center}
\begin{threeparttable}
\centering \footnotesize
\caption{\footnotesize Risk Premia}\label{tab:RP_2cluster}
\begin{tabular}{L{0.15\textwidth} C{0.23\textwidth}C{0.23\textwidth}C{0.23\textwidth}}
\toprule
\multicolumn{3}{l}{\textbf{Panel A:} Bitcoin Premium} \\
\bottomrule
                          & Overall & HV & LV \\
\cmidrule(r){2-4}
$\widehat{\operatorname{BP}}=\widehat{\mu}_{\p} - \widehat{\mu}_{\q}$ &   {\textbf{0.66}} &  {\textbf{0.69}}  &  {\textbf{0.62}} 
\\ 
$\widehat{\mu}_{\p}$ &   {0.67} &  {0.70}  &  {0.63} \\
$\widehat{\mu}_{\q}$ &  0.01 & 0.01  &  0.01 \\   
\bottomrule
\multicolumn{3}{l}{\textbf{Panel B:} Bitcoin Variance Risk Premium} \\
\bottomrule
                          & Overall & HV & LV \\
\cmidrule(r){2-4}
$\widehat{\operatorname{VRP}}=\widehat{\sigma}^2_\q -\widehat{\sigma}^2_\p$   &   {\textbf{0.14}} & {\textbf{0.12}}  &  {\textbf{0.17}\tmark[{\makebox[0pt][l]{}}]} \\ 
$\widehat{\sigma}^2_\q$   &  0.72 & 0.86\tmark[{\makebox[0pt][l]{***}}]  &  0.46\tmark[{\makebox[0pt][l]{***}}] \\ 
$\widehat{\sigma}^2_\p$   &   0.58 & 0.74\tmark[{\makebox[0pt][l]{***}}]  &   0.29\tmark[{\makebox[0pt][l]{***}}] \\ 
Days    &  1017 & 649      &  368  \\ 
\bottomrule
\end{tabular}
\renewcommand{\baselinestretch}{0.8}\footnotesize
\begin{minipage}{0.93\textwidth}
\textbf{Panel A:} Estimates of the unconditional BP and conditional $\bp_i$. 
\textbf{Panel B:} Estimates of the unconditional BVRP and conditional $\bvrp_i$. Unconditional estimates are referred to as 'Overall', and the conditional ones are cluster-specific for $i \in \{\text{HV},\text{LV}\}$. $\widehat{\mu}_{\p}=\frac{365}{27} \int_{-1}^{1}x\{\hat{p}(x)\}dx$ and $\hat{\sigma}^2_{\p}(R)$ is the sample mean of realized variances. $\mu_{\q}$ is estimated using 1-month Treasury bill rates and for $\hat{\sigma}^2_{\q}(R)$ we use square of BVIX. All the estimates are annualized. ANOVA is applied to test whether the conditional estimates are different than the unconditional ones ($H_0:$ no difference), with $1\%(^{***})$, $5\%(^{**})$ and $10\%(^{*})$ denoting significance level. 
\end{minipage}
\end{threeparttable}
\end{center}

\begin{figure}[tbp]
    \centering
    \begin{subfigure}[b]{0.49\textwidth}
        \centering
        \includegraphics[width=\textwidth]{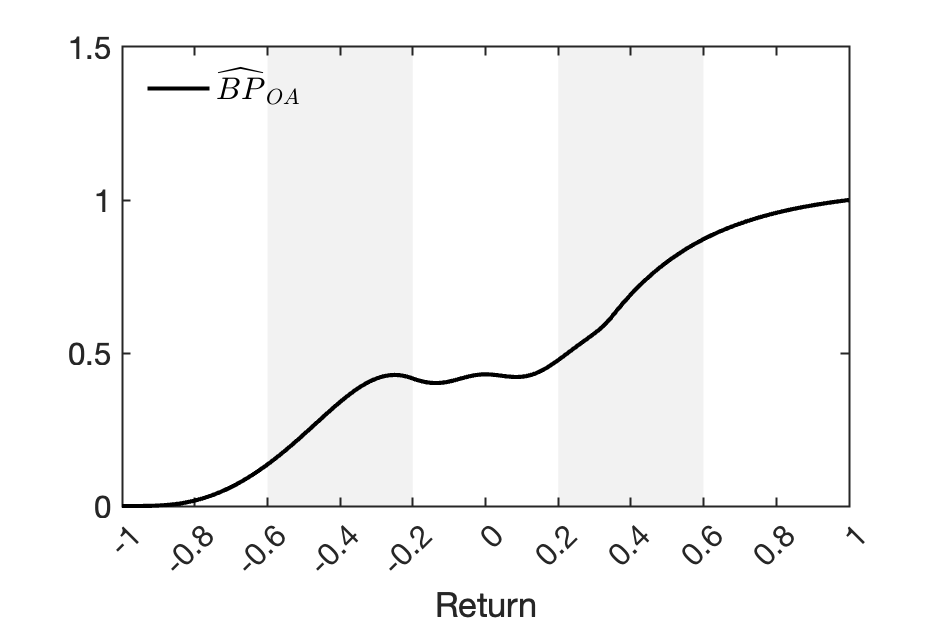}
    \end{subfigure}
    \hfill
    \begin{subfigure}[b]{0.49\textwidth}
        \centering
        \includegraphics[width=\textwidth]{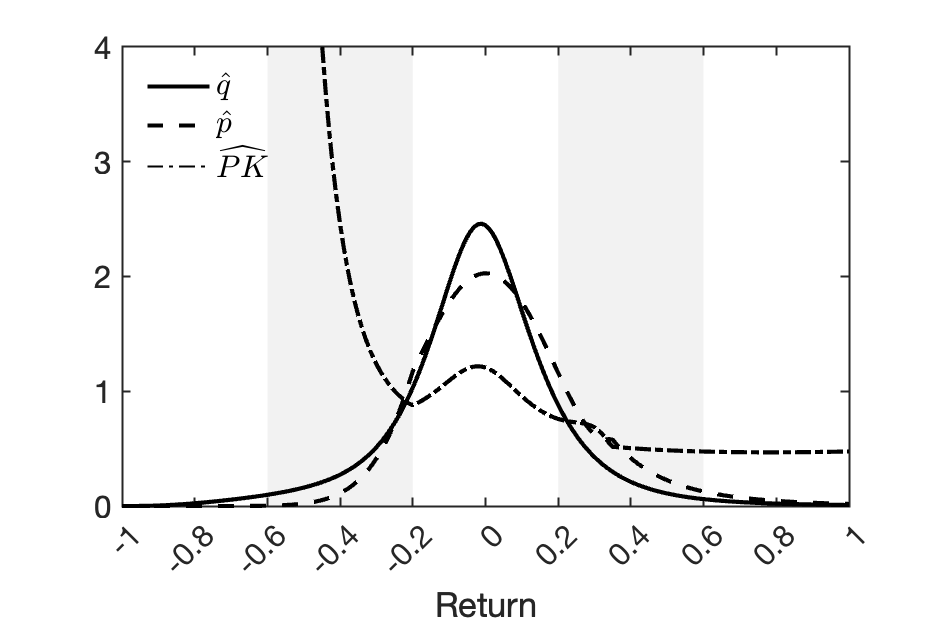}
    \end{subfigure}
    \begin{subfigure}[b]{0.49\textwidth}
        \centering
        \includegraphics[width=\textwidth]{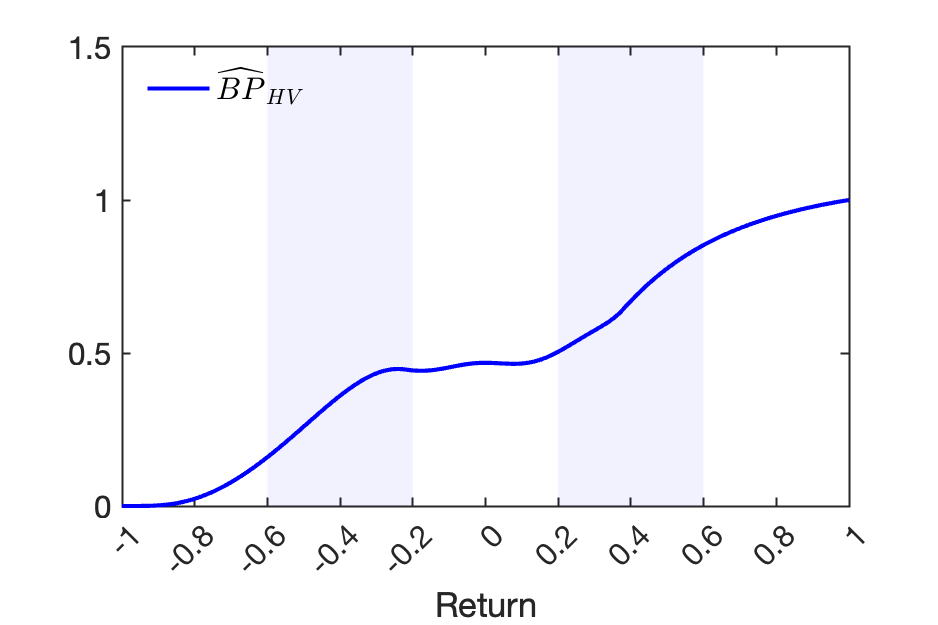}
    \end{subfigure}
    \hfill
    \begin{subfigure}[b]{0.49\textwidth}
        \centering
        \includegraphics[width=\textwidth]{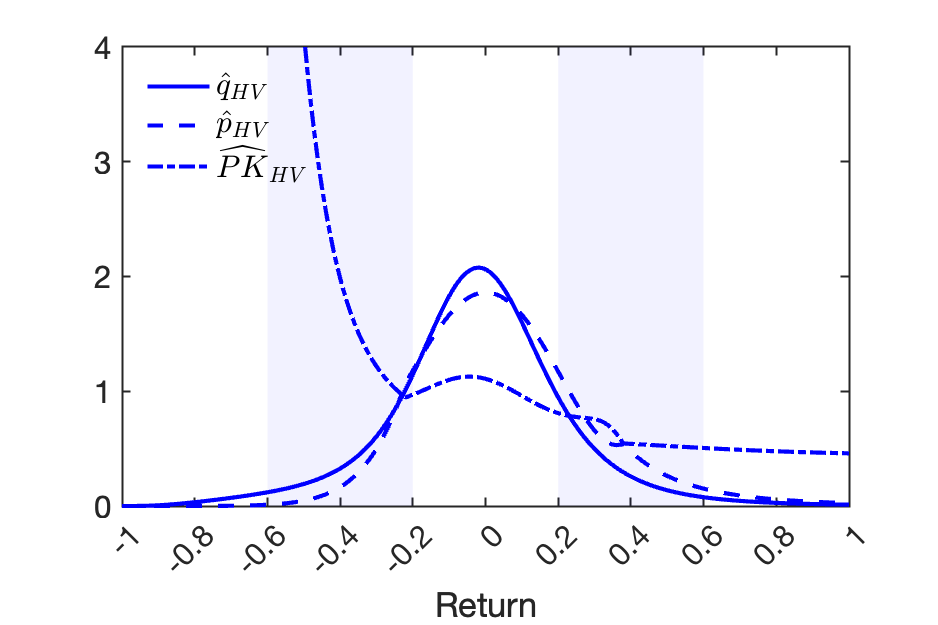}
    \end{subfigure}
    \begin{subfigure}[b]{0.49\textwidth}
        \centering
        \includegraphics[width=\textwidth]{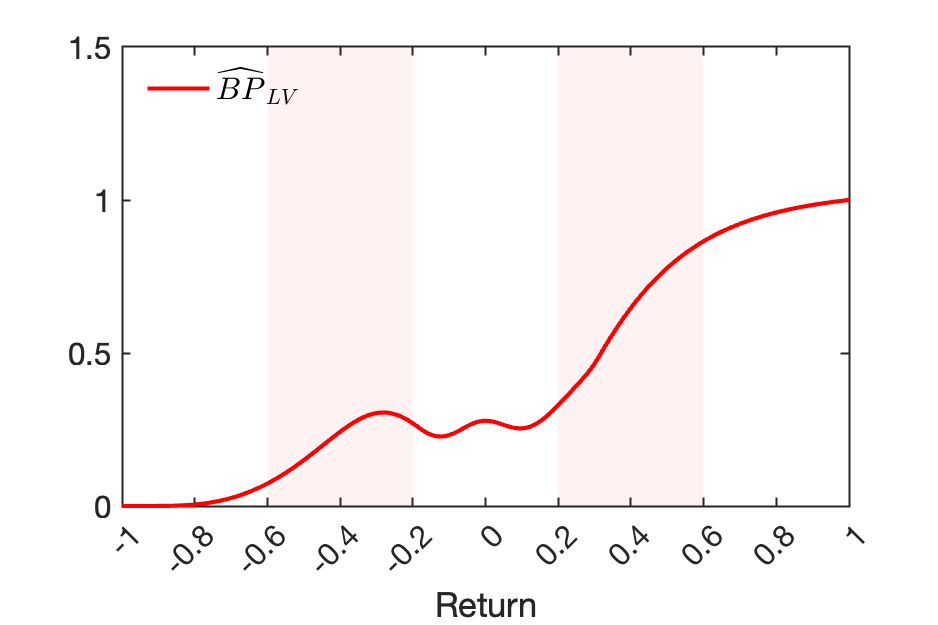}
    \end{subfigure}
    \hfill
    \begin{subfigure}[b]{0.49\textwidth}
        \centering
        \includegraphics[width=\textwidth]{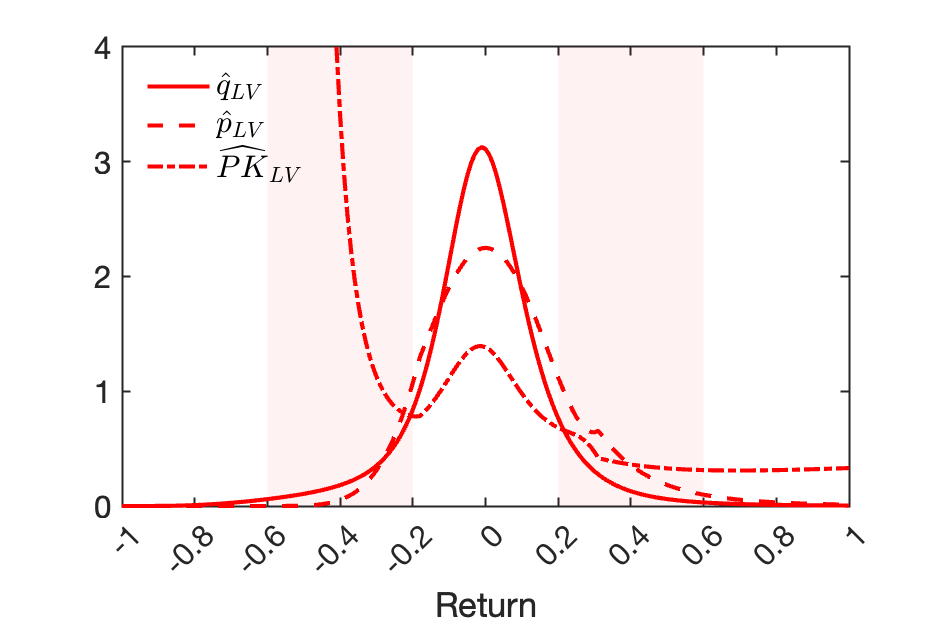}
    \end{subfigure}
\caption{First column includes estimated BP for overall (black), HV cluster (blue), and the LV cluster (red). The second column includes estimated PK, physical, and risk-neutral density for overall (black), HV cluster (blue) and the LV cluster (red). The shaded areas mark the returns range [-0.6, -0.2] and [0.2, 0.6]. \textbf{First row.} For the overall sample, the shaded areas contribute  {28.53}\% and  {38.66}\% to the overall BP, respectively. \textbf{Second row.} For the HV regime, the shaded area contributes  {28.47}\% and {34.21}\% to the BP. \textbf{Third row.} The shaded areas of the LV cluster contribute  {20.44}\% and  {52.30}\% to the BP.}\label{fig:all_res}  
\end{figure}

The unconditional decomposition of the $\bp(r)$ exhibited in Figure \ref{fig:all_res} reveals that $\bp(r)$ is mostly increasing for return states smaller than -20\% and larger than 20\%, reflecting that these states contribute positively to the Bitcoin premium.  BP attributable to positive and negative returns has a similar share, with positive returns contributing slightly more to the BP. 
For illustration, we focus on return states that roughly span from one to three standard deviations away from the mean.\footnote{For the Bitcoin Index, the standard deviation is 0.22 for the unconditional physical distribution of monthly returns. By contrast, the results in \citet{Beason2022-ht} suggest that the standard deviation for S\&P 500 monthly returns is 0.04.} We find that negative monthly returns between -60\% and -20\% and positive returns ranging from 20\% to 60\% contribute $28.53 \%$ and $38.66 \%$, respectively of the BP. The comparable attribution of the Bitcoin premium between positive and negative returns contrasts with the results in the S\&P 500 Index market, where 80\% of EP is attributable to monthly returns below -10\%, as reported by \citet{Beason2022-ht}, while positive returns have a much smaller (positive) share.\footnote{\citet{Beason2022-ht} primarily focus on negative returns and do not quantify the risk premia associated with positive returns. From their Figure 1, we can determine that the positive equity premium is linked to returns between 0\% and 5\%, and accounts for approximately 25\% of the overall equity premium. In addition, for S\&P 500 returns between 5\% and 20\%, the equity premium is decreasing, and these return states contribute negatively to the first moment premium.} The increasing $\bp(r)$ for positive returns indicates that OTM calls are profitable on average, as investors require a positive premium for holding them. Conversely, the increasing $\bp(r)$ for negative returns indicates that OTM puts are expensive, as investors are paying a premium to hold them for hedging purposes. This interpretation is supported by the shape and values of the pricing kernel.\footnote{PK takes values above one for returns below -20\%, and OTM puts behave like hedge assets, and below one for returns above 20\%, and OTM calls are considered risky assets.} The $\pk(r)$ is generally decreasing, except for a segment of Bitcoin returns between -20\% and 0\%, which creates a hump. The hump is above one, indicating that ATM options are expensive, leading to negative returns for both ATM put and call options. This also highlights the presence of a variance risk premium.\footnote{Expensive ATM options have been used in the literature to document the variance risk premium (\cite{Coval2001-cd}).}

Next, we examine closely the risk functions across different market conditions. Figure \ref{fig:all_res} also displays the $\bp(r)$  for the two clusters, revealing several noteworthy characteristics. Notably, the shape of $\bp(r)$ in the HV cluster is more akin to the unconditional $\bp(r)$. 
But there are also some important differences. 
The BP attributable to negative returns remains relatively constant, with a share of 28.47\% for returns between -60\% and -20\%, while the impact of positive returns is slightly diminished, with returns from 20\% to 60\% contributing only 34.21\% to the BP.
In the LV cluster, $\bp(r)$  exhibits a substantial increase in the region of positive returns. We identify a novel pattern for returns ranging from 20\% to 60\% that exhibit a significant {52.30\%} positive contribution to the BP. 
In contrast, the share of BP attributable to negative returns ranging from -60\% to -20\% is only 20.44\%. The BP attributable to returns below -60\% and above 60\% is relatively small. 
These results suggest that large positive returns, but not extreme returns, are the main source of Bitcoin risk premia during calm markets. Investors are more concerned with the risk compensation for the upside risk and less concerned with hedging the downside risk. 

The pricing kernel $\pk(r)$ is hump-shaped in both market regimes, with the hump more pronounced in the LV regime. Higher values (above one) of the PK around ATM are consistent with a higher BVRP during less volatile markets.\footnote{The higher hump of the PK in the low-volatility regime may suggest increased activity from volatility traders who are hedging against volatility risk. This interpretation is supported by \citet{Alexander2023-wq}, who states that high net buying pressure in ATM options, when tied to rising ATM implied volatility, is a strong indicator of volatility traders.} Furthermore, Figure \ref{fig:all_res} shows that $\pk(r)$ is considerably steeper and takes higher values for large negative returns below -30\% when volatility is low. This result reinforces the finding of \citet{Schreindorfer2025-fa} that negative returns are substantially more painful to investors in periods of low volatility. It is important to note, however, that while \citet{Schreindorfer2025-fa} employ a parametric specification of the pricing kernel, we use a fully nonparametric approach and obtain the same results.\footnote{In the appendix, the authors display the parametric estimates of the PK for the entire range of returns, yet they do not delve into an extensive explanation of the pricing kernel variation for the positive range of returns. Their pricing kernel is U-shaped, with a steeper increasing region for the positive returns. Our non-parametric estimates, in contrast, show a higher pricing kernel for this region in volatile markets.} Drawing on studies that identify and dissect the variance risk premium by looking at the entire range of returns (e.g., \cite{Kilic2019-om} and \cite{Almeida2024-dt}), we can also establish that while most of the BP is attributable to positive returns, the negative returns contribute mostly to the BVRP.\footnote{Focusing on near-to-expiry options, \citet{Almeida2024-dt} show that a positive VRP occurs when PK exhibits a U-shape and deep-out-of-the-money (DOTM) call options yield negative returns.} 
Focusing on the shape variation of $\pk(r)$, our analysis also reveals a positive relationship between the height of the peak and the BVRP. This is different from the result reported by \citet{Grith2017-dj}, who find a hump-shaped pricing kernel when VRP is low and a U-shaped pricing kernel when it is high, providing additional support for the potential disconnect between variance and variance risk premium. 

The results in Table \ref{tab:price_of_risk} show that the probabilities of large price movements away from the current value, from one to three standard deviations, display asymmetries between the left and right sides of the distribution for both conditional and unconditional estimates. In particular, the probability of an increase in this segment is, on average, twice as large as the probability of a decrease. Risk compensation on the left and right segments is balanced for the unconditional and the turbulent markets. In contrast, during less volatile markets, the main source of risk premia for Bitcoin is large positive returns. Their probability is relatively small (around 16\%) in this regime. These results show that investors are primarily rewarded for less frequent, large positive returns during less volatile markets.

\begin{center}
\begin{threeparttable}
\centering \footnotesize
\caption{\footnotesize Characteristics of BP, $\q$ and $\p$ in influential states}\label{tab:price_of_risk}
\begin{tabular}{L{0.06\textwidth} C{0.16\textwidth}C{0.08\textwidth}C{0.1\textwidth}C{0.15\textwidth}C{0.08\textwidth}C{0.10\textwidth}}
\toprule
& \multicolumn{3}{c}{Negative states} & \multicolumn{3}{c}{Positive states} \\
\cmidrule(r){2-4}\cmidrule(r){5-7}
                          & BP(-0.2)-BP(-0.6) & $\int_{-0.6}^{-0.2}p(r)dr$ & $\frac{\int_{-0.6}^{-0.2}q(r)dr}{\int_{-0.6}^{-0.2}p(r)dr}$ & BP(0.2)-BP(0.6) & $\int_{0.2}^{0.6}p(r)dr$ & $\frac{\int_{0.2}^{0.6}q(r)dr}{\int_{0.2}^{0.6}p(r)dr}$ \\
\cmidrule(r){2-4}\cmidrule(r){5-7}
Overall & 0.29 & 0.09 & 1.45 & 0.39 & 0.18 & 0.61 \\ 
HV      & 0.28 & 0.11 & 1.40 & 0.34 & 0.19 & 0.67 \\
LV      & 0.20 & 0.07 & 1.33 & 0.52 & 0.16 & 0.47 \\  
\bottomrule
\end{tabular}
\renewcommand{\baselinestretch}{0.8}\footnotesize
\begin{minipage}{0.93\textwidth}
BP(-0.2)-BP(-0.6) and BP(0.2)-BP(0.6) are BP contributions on the intervals. $\int p(r)dr$ is the physical probability on such states and $\frac{\int q(r)dr}{\int p(r)dr}$ is the corresponding risk price.
\end{minipage}
\end{threeparttable}
\end{center}

In Table \ref{tab:price_of_risk} we also calculate the price of risk as the ratio of the average risk-neutral density to the physical density, as in \citet{Beason2022-ht}, which allows us to draw a comparison to the S\&P 500 market.
For negative states between -60\% and -20\%, the risk price for Bitcoin is approximately 1.45, which is lower than the 2.63 for the S\&P 500 as reported by \citet{Beason2022-ht}  for returns -30\% and -10\%, but comparable to levels found in \citet{Campbell1999-de}, \citet{Bansal2004-qr}, \citet{Barro2009-yp}, and \citet{Wachter2013-qj}.\footnote{A direct comparison on the same segment -30\% and -10\% would produces even lower prices for risk for the Bitcoin market on this segment. Similarly, calculating the price of risk for S\&P 500 for segments beyond -30\% would result in higher risk prices for large returns.} These results show that the Arrow-Debreu prices are smaller on average in the Bitcoin market than in the S\&P 500 market, as investors are willing to pay less for hedging the downside risk. Our analysis extends to the positive returns and finds a price of risk of 0.61 for returns between 20\% and 60\%; a direct comparison with the above-mentioned studies is not available.  


\begin{figure}[htbp]
  \centering
  \subfloat[Lower bounds of BP]{\includegraphics[width=0.5\textwidth]{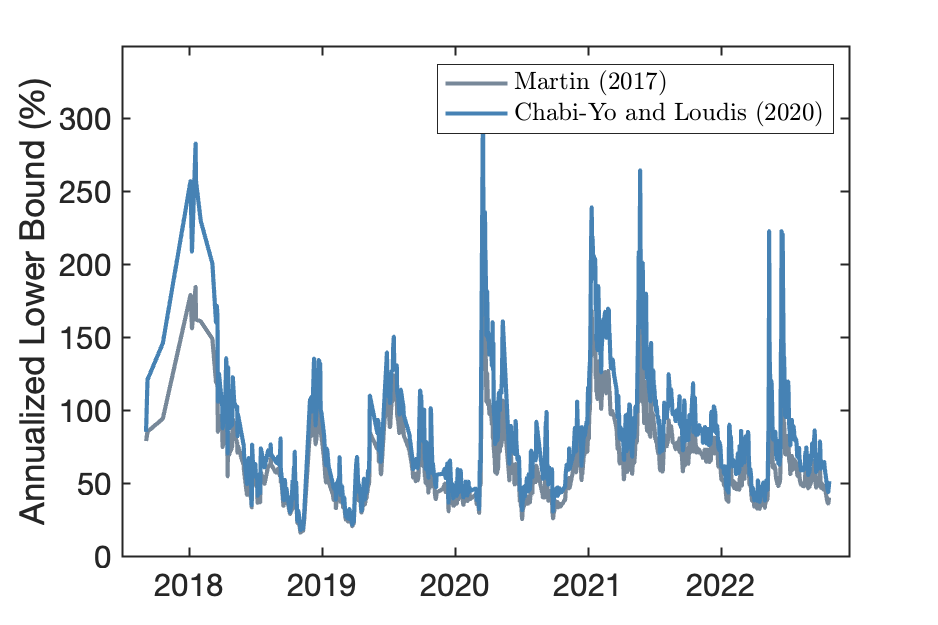}}
  \hfill
  \subfloat[VRP]{\includegraphics[width=0.5\textwidth]{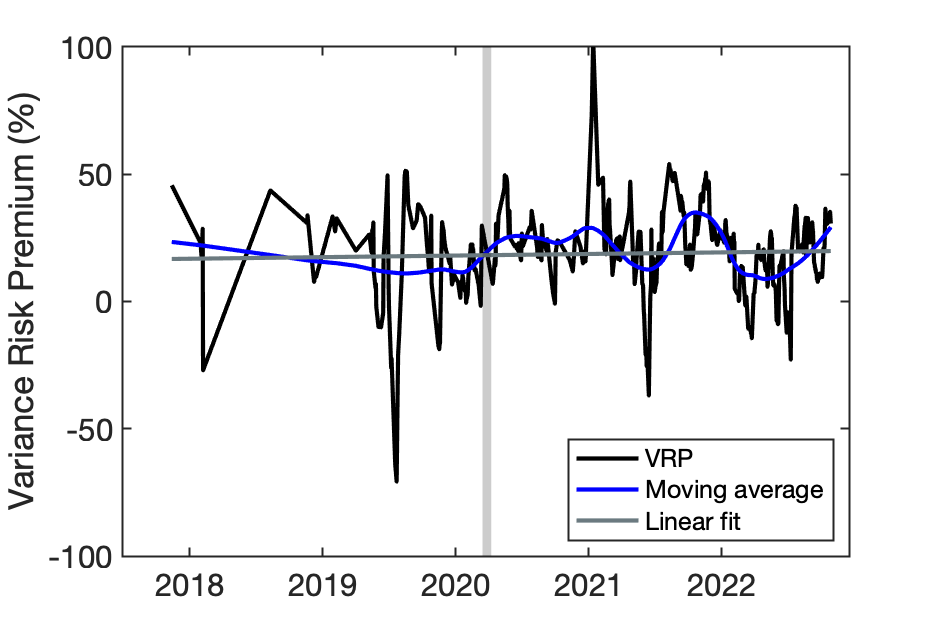}}
      \hfill
  \subfloat[Bitcoin Index]{\includegraphics[width=0.5\textwidth]{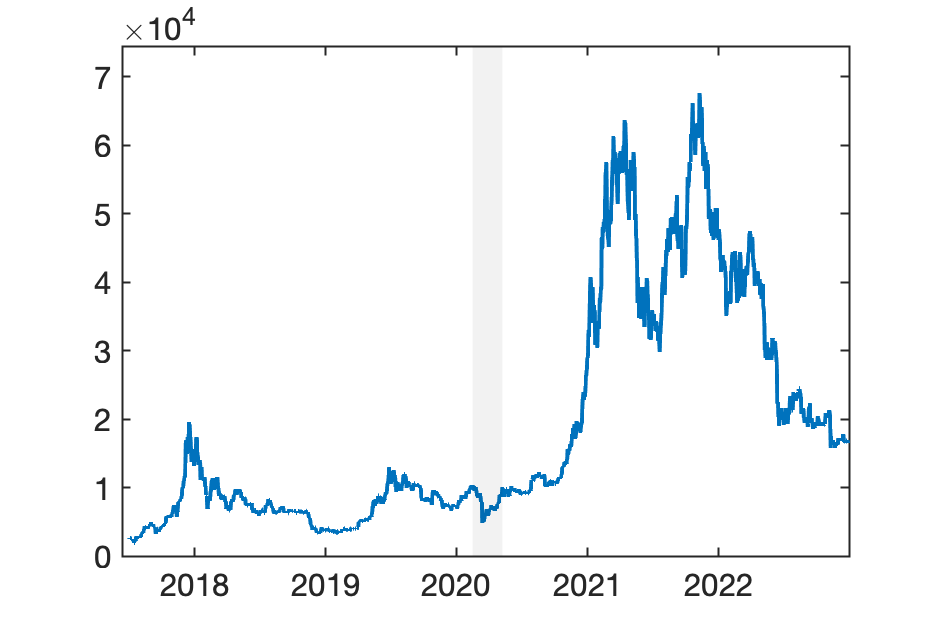}}
  \hfill
  \subfloat[$\text{BVIX}^2$ and $\rv$]{\includegraphics[width=0.5\textwidth]{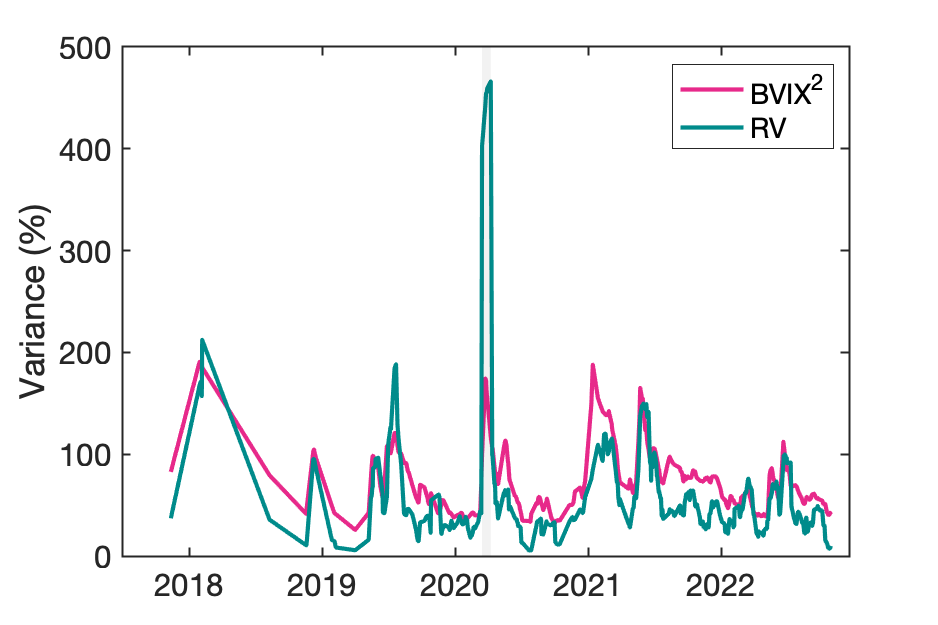}}
  \caption{BP lower bounds, VRP, Bitcoin Index, $\text{BVIX}^2$ and $\rv$ between July 2017 and December 2022. 
  (a) The calculation of lower bounds follows the methodologies of \citet{Martin2017-sl} and \citet{Chabi-Yo2020-tv} based on the empirical risk-neutral density. 
  The conditional lower bound risk premia range from 20\% to 200\%. The average lower bounds calculated using the methods of Martin (2017) and Chabi-Yo and Loudis (2020) are 65.65\% and 84.78\%, respectively. (b) $\bvrp_t$ is calculated as the difference between $\text{BVIX}_t^2$ and $\rv_t$. $\bvrp_t$ estimates less than -2 are excluded, and the corresponding days are highlighted with shaded areas. Trends are illustrated using moving averages and linear fit, respectively.
  (c)  The shaded area marks the sudden fall and rebound of the Bitcoin Index between Mar 15, 2020, and April 8, 2020. (d) Generally, $\text{BVIX}^2_t$ remains above $\rv_t$, except for about a month in 2020 when $\rv$ approached 500\% as Bitcoin Index surprisingly fell from \$10,000 to \$5,000 and subsequently rebounded.}
\label{fig:BPLB_VRP_Bitcoin_BVIX_RV_overtime}
\end{figure}

In Figure \ref{fig:BPLB_VRP_Bitcoin_BVIX_RV_overtime} we display the BP and VRP estimates over time. These estimates are noisy but convey a few key features. Figure \ref{fig:BPLB_VRP_Bitcoin_BVIX_RV_overtime} (a) displays the time-varying lower bounds of the Bitcoin premium estimated using methods of \citet{Martin2017-sl, Chabi-Yo2020-tv}. These bounds are high, especially during market disruptions.  
At their peak, the annualized lower bounds exceeded 200\%, indicating extremely high required compensation for holding Bitcoin. These correspond to significant changes in Bitcoin's price during the early 2018 crash, the March 2020 COVID-19 crash, and the 2021 regulatory news and macro events. During calm periods, e.g., 2019, the lower bounds fall to around 50-100\%, which is still higher than traditional assets. 
Figure \ref{fig:BPLB_VRP_Bitcoin_BVIX_RV_overtime} (a) and (d) show that as the lower bounds increase, the variances also rise. In contrast, the BVRP series in \ref{fig:BPLB_VRP_Bitcoin_BVIX_RV_overtime} (b) highlights the alternating sign movements between the spikes of the first moment lower bound and the BVRP.\footnote{In general, $\text{BVIX}^2$ and $\rv$ series indicate a tendency of positive comovement, with $\text{BVIX}^2$ exceeding $\rv$, except from a period between March 15, 2020 and April 8, 2020 when $\rv$ significantly surpassed $\text{BVIX}^2$ and VRP takes values below -2 as marked by the shaded area in Figure \ref{fig:BPLB_VRP_Bitcoin_BVIX_RV_overtime} (b). However, their difference does not always move in the same direction with the volatility. }

\section{Conclusion} \label{sec:conclusion}
This work estimates the Bitcoin Index's return premium and volatility risk premium using joint options and returns data over the most extensive available period. A Bitcoin premium decomposition is applied as a function of returns. We also propose a new functional clustering method applied to a sequence of time series of Bitcoin risk-neutral measures that allows us to obtain conditional measures for Bitcoin's first moment and variance risk premia. Overall, we find that Bitcoin's first and second moment risk premia, and the premium attributable to positive returns are all much larger than the corresponding measures for traditional assets like the S\&P 500. We also find significant variation in these metrics between low- and high-volatility regimes, suggesting that volatility is an important state variable driving risk in the Bitcoin market. 

\printbibliography

\end{refsection}

\clearpage

\begin{refsection}
\newpage
\appendix

\DoToC

\newpage
\section{Main Appendix}

\setcounter{figure}{0}
\renewcommand{\thefigure}{A\arabic{figure}}
\setcounter{table}{0}
\renewcommand{\thetable}{A\arabic{table}}
\subsection{Data Analysis} \label{app:data_analysis}



Two types of BTC options are traded on Deribit: those with shorter tenors--up to two days--that expire daily at 08:00 UTC and those with longer tenors that expire on Fridays at 08:00 UTC--more than two days. We utilize the data from the latter options in our study.

\begin{figure}[H]
  \centering
  \begin{tabular}{cc}
    \includegraphics[width=0.9\textwidth]{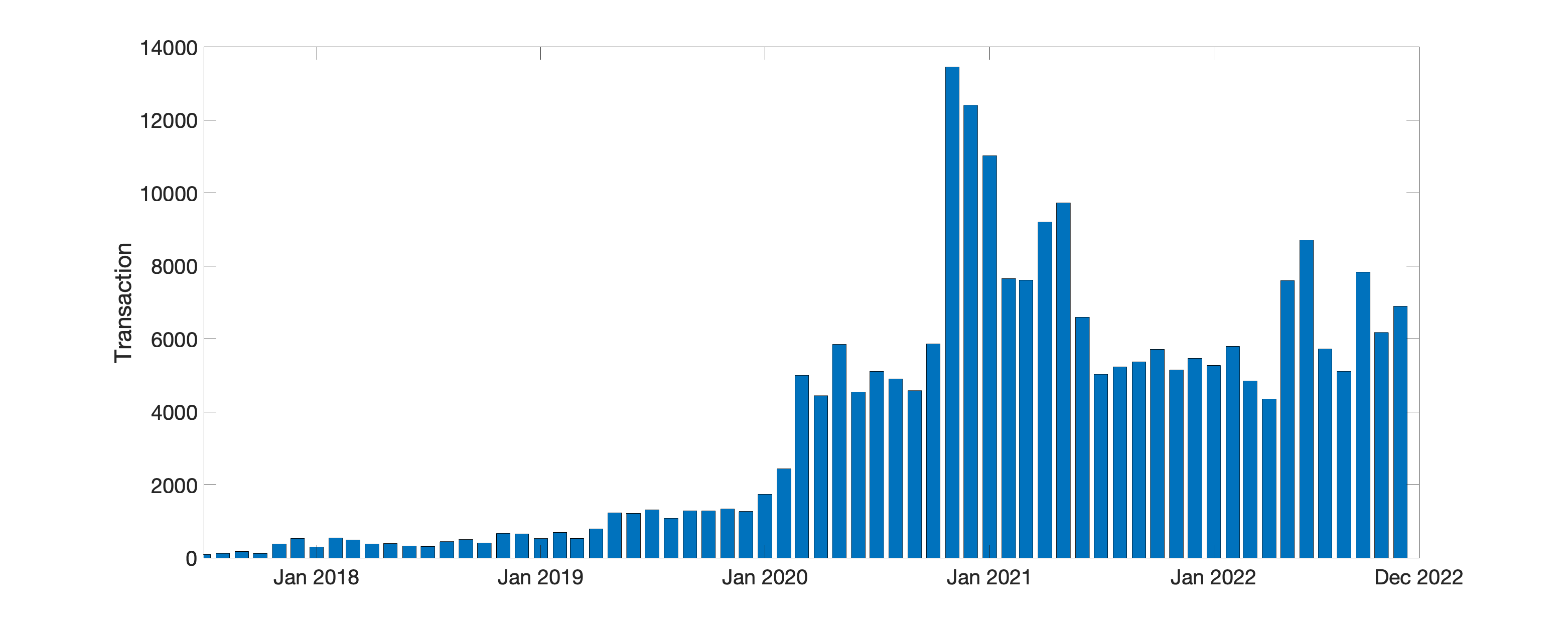} \\
  \end{tabular}
  \caption{Average daily BTC option transaction per month}\label{fig:transaction_daily}  
\end{figure}

Figure \ref{fig:transaction_daily} illustrates the average number of BTC option transactions per month.

Table \ref{tab:IV} gives an overview of the average implied volatility on different batches. We notice for both call and put options, IV initially decreases as moneyness increases and then rises past ATM, representing a "volatility smile" commonly seen in the traditional security markets. Furthermore, options with shorter maturity, particularly those deep OTM and deep ITM, tend to exhibit higher levels of IV. Notably, put options generally display higher IV compared to call options.

\begin{center}
\begin{threeparttable}
\centering \footnotesize
\caption{\footnotesize  Implied volatility of BTC options [in level]}\label{tab:IV}
\begin{tabular}{L{0.1\textwidth} C{0.1\textwidth}C{0.1\textwidth}C{0.1\textwidth}C{0.1\textwidth}C{0.1\textwidth}}
\toprule
        & \multicolumn{5}{c}{Call options} \\
\cmidrule(r){2-6}
Moneyness & (0, 9] & [10, 26] & [27, 33] & >33 & Average\\
\cmidrule(r){1-6}
$<-0.6 $     &      4.74 &      2.70 &     1.76 &     1.26 &      1.45 \\ 
             &      (120)&     (220) &     (130)&   (3,647)&    (4,117)\\ 
$[-0.6,-0.2)$&      1.82 &      1.26 &     1.03 &     0.95 &      1.14 \\ 
             &    (7,474)&   (7,546) &   (2,850)&  (29,628)&   (47,498)\\ 
$[-0.2,0.2]$ &      0.78 &      0.75 &     0.75 &     0.78 &      0.77 \\ 
             &(1,914,619)& (593,931) &  (91,586)& (328,960)&(2,929,096)\\ 
$(0.2,0.6]$  &      1.23 &      0.94 &     0.86 &     0.83 &      0.92 \\ 
             &   (87,258)& (192,329) &  (49,814)& (291,264)&  (620,665)\\ 
$>0.6$       &      1.99 &      1.38 &     1.17 &     0.98 &      1.04 \\ 
             &    (4,977)&  (26,024) &  (13,965)& (294,199)&  (339,165)\\ 
Average      &      0.81 &      0.82 &     0.83 &     0.86 &      0.82 \\ 
             &(2,014,448)& (820,050) & (158,345)& (947,698)&(3,940,541)\\ 
\midrule
& \multicolumn{5}{c}{Put options}\\
\cmidrule(r){2-6}
Moneyness & (0, 9] & [10, 26] & [27, 33] & >33 & Average\\
\cmidrule(r){1-6}
$<-0.6 $     &      3.70 &      2.13 &     1.85 &     1.29 &      1.35 \\  
             &       (69)&   (1,554) &   (1,429)&  (34,658)&   (37,710)\\
$[-0.6,-0.2)$&      1.54 &      1.18 &     1.05 &     0.95 &      1.12 \\
             &   (97,793)& (172,290) &  (39,570)& (277,739)&  (587,392)\\  
$[-0.2,0.2]$ &      0.84 &      0.80 &     0.77 &     0.81 &      0.83 \\ 
             &(1,776,761)& (568,189) &  (82,503)& (352,108)&(2,779,561)\\ 
$(0.2,0.6]$  &      2.31 &      1.26 &     0.81 &     0.93 &      1.35 \\ 
             &   (12,187)&  (11,404) &   (2,665)&  (22,298)&   (48,554)\\ 
$>0.6$       &      3.15 &      1.92 &     1.40 &     1.05 &      1.36 \\ 
             &      (922)&   (2,642) &     (904)&  (10,335)&   (14,803)\\ 
Average      &      0.89 &      0.90 &     0.88 &     0.90 &      0.89 \\ 
             &(1,887,732)& (756,079) & (127,071)& (997,138)&(3,468,020)\\ 
\bottomrule
\end{tabular}
\renewcommand{\baselinestretch}{0.8}\footnotesize
\begin{minipage}{0.75\textwidth}
This table presents the average implied volatility of BTC options across moneyness and maturity. 
The columns are categorized based on the time to maturity measured in days. The IVs are sourced from Deribit. The numbers of observations is provided in parentheses. 
\end{minipage}
\end{threeparttable}
\end{center}

Table \ref{tab:summary_transaction} presents the transaction patterns of call and put options, classified into different moneyness and maturity groups. The results reveal that OTM options are predominant for both call and put options, accounting for more than 60\% of the total, with deep OTM options making up more than 35\%. In contrast, in-the-money (ITM) options constitute less than 10\%, with deep ITM options accounting for less than 4\%. Regarding the term structure, more than half of the options have maturities of less than 10 days, with a slightly higher proportion of put options (54.43\%) compared to call options (51.12\%). Moreover, call options with maturities of more than 33 days constitute 24.05\%, whereas put options with maturities of more than 33 days account for 20.10\%.

\begin{center}
\begin{threeparttable}
\centering \footnotesize
\caption{\footnotesize Summary statistics on option contracts of BTC options [in \%]}\label{tab:summary_transaction}
\begin{tabular}{L{0.1\textwidth} C{0.1\textwidth}C{0.1\textwidth}C{0.1\textwidth}C{0.1\textwidth}C{0.1\textwidth}}
\toprule
        & \multicolumn{5}{c}{Call options} \\
\cmidrule(r){2-6}
Moneyness & (0, 9] & [10, 26] & [27, 33] & >33 & Subtotal\\
\cmidrule(r){1-6}
$<-0.6 $      &\;\;0.00&\;\;0.01 &    0.00 &\;\;0.09 &\;\;\;0.10 \\ 
$[-0.6,-0.2)$ &\;\;0.19&\;\;0.19 &    0.07 &\;\;0.75 &\;\;\;1.21 \\ 
$[-0.2,0.2]$  &   48.59&   15.07 &    2.32 &\;\;8.35 &   \;74.33 \\ 
$(0.2,0.6]$   &\;\;2.21&\;\;4.88 &    1.26 &\;\;7.39 &   \;15.75 \\ 
$>0.6$        &\;\;0.13&\;\;0.66 &    0.35 &\;\;7.47 &\;\;\;8.61 \\ 
Total         &   51.12&   20.81 &    4.02 &   24.05 &    100.00 \\
\midrule
& \multicolumn{5}{c}{Put options}\\
\cmidrule(r){2-6}
$<-0.6)$      &\;\;0.02 &\;\;0.04 & 0.04 &\;\;1.00 &\;\;\;1.09 \\ 
$[-0.6,-0.2)$ &\;\;2.82 &\;\;4.97 & 1.14 &\;\;8.01 &   \;16.94 \\ 
$[-0.2,0.2]$  &   51.23 &   16.38 & 2.38 &   10.15 &   \;80.15 \\ 
$(0.2,0.6]$   &\;\;0.35 &\;\;0.33 & 0.08 &\;\;0.64 &\;\;\;1.40 \\ 
$>0.6$        &\;\;0.03 &\;\;0.08 & 0.03 &\;\;0.30 &\;\;\;0.43 \\ 
Total         &   54.43 &   21.80 & 3.66 &   20.10 &    100.00 \\
\bottomrule
\end{tabular}
\renewcommand{\baselinestretch}{0.8}\footnotesize
\begin{minipage}{0.75\textwidth}
This table presents the proportion of traded BTC option contracts over moneyness and maturity. The sample covers transactions between July 1, 2017 and December 17, 2022. 
The columns are categorized based on the time to maturity in days. The transactions are measured as the number of traded contracts. 
\end{minipage}
\end{threeparttable}
\end{center}

Table \ref{tab:summary_quantity} provides summary statistics on option quantity in BTC units, given that each option is denominated in BTC. The distribution of quantity closely mirrors that of transaction contracts, with an even greater proportion of out-of-the-money options. Table \ref{tab:summary_volume} presents the summary statistics on option transaction volume in USD, calculated as the traded quantity multiplied by the option price in USD. Options with longer maturities and in-the-money options typically possess higher prices, resulting in over half of the total volume being attributed to long-maturity options. Additionally, the OTM volume portion is lower than transaction and quantity due to their lower prices. 

\begin{center}
\begin{threeparttable}
\centering \footnotesize
\caption{\footnotesize  Summary statistics on BTC option volume [in \%]}\label{tab:summary_quantity}
\begin{tabular}{L{0.1\textwidth} C{0.1\textwidth}C{0.1\textwidth}C{0.1\textwidth}C{0.1\textwidth}C{0.1\textwidth}}
\toprule
        & \multicolumn{5}{c}{Call options} \\
\cmidrule(r){2-6}
Moneyness & (0, 9] & [10, 26] & [27, 33] & >33 & Subtotal\\
\cmidrule(r){2-6}
$<-0.6 $      &\;\;0.00 &\;\;0.00 & 0.00 &\;\;0.07 &\;\;\;0.07 \\ 
$[-0.6,-0.2)$ &\;\;0.13 &\;\;0.14 & 0.03 &\;\;0.47 &\;\;\;0.77 \\ 
$[-0.2,0.2]$  &   41.35 &   17.09 & 2.81 &\;\;8.14 & \;\;69.40 \\ 
$(0.2,0.6]$   &\;\;2.56 &\;\;5.80 & 1.82 &\;\;9.18 & \;\;19.36 \\ 
$>0.6$        &\;\;0.15 &\;\;0.95 & 0.41 &\;\;8.89 & \;\;10.40 \\ 
Total         &   44.19 &   23.98 & 5.07 &   26.75 &    100.00 \\
\midrule
& \multicolumn{5}{c}{Put options}\\
\cmidrule(r){2-6}
$<-0.6)$      &\;\;0.00 &\;\;0.04 & 0.03 &\;\;0.84 &\;\;\;0.91 \\ 
$[-0.6,-0.2)$ &\;\;4.09 &\;\;5.83 & 1.50 &\;\;8.85 &   \;20.28 \\ 
$[-0.2,0.2]$  &   46.47 &   19.21 & 2.82 &\;\;9.28 &   \;77.78 \\ 
$(0.2,0.6]$   &\;\;0.15 &\;\;0.24 & 0.04 &\;\;0.39 &\;\;\;0.82 \\ 
$>0.6$        &\;\;0.02 &\;\;0.03 & 0.00 &\;\;0.16 &\;\;\;0.22 \\ 
Total         &   50.74 &   25.35 & 4.40 &   19.51 &    100.00 \\
\bottomrule
\end{tabular}
\renewcommand{\baselinestretch}{0.8}\footnotesize
\begin{minipage}{0.75\textwidth}
This table presents the proportion of volume [in \%] of the BTC option data over moneyness and maturity. The data spans from July 1, 2017, to December 17, 2022. The columns are categorized based on the time to maturity in days. The volume is measured in terms of the number of BTC units. 
\end{minipage}
\end{threeparttable}
\end{center}

\begin{center}
\begin{threeparttable}
\centering \footnotesize
\caption{\footnotesize  Summary statistics on BTC option transaction value valued in USD [in \%]}\label{tab:summary_volume}
\begin{tabular}{L{0.1\textwidth} C{0.1\textwidth}C{0.1\textwidth}C{0.1\textwidth}C{0.1\textwidth}C{0.1\textwidth}}
\toprule
        & \multicolumn{5}{c}{Call options} \\
\cmidrule(r){2-6}
Moneyness & (0, 9] & [10, 26] & [27, 33] & >33 & Subtotal\\
\cmidrule(r){1-6}
$<-0.6 $      &\;\;0.03 &\;\;0.06 & 0.03 &\;\;1.77 &\;\;\;1.89 \\ 
$[-0.6,-0.2)$ &\;\;0.88 &\;\;1.54 & 0.24 &\;\;4.88 &\;\;\;7.54 \\ 
$[-0.2,0.2]$  &   17.57 &   18.90 & 4.31 &   23.69 & \;\;64.48 \\ 
$(0.2,0.6]$   &\;\;0.26 &\;\;2.07 & 0.95 &   14.14 & \;\;17.41 \\ 
$>0.6$        &\;\;0.01 &\;\;0.24 & 0.07 &\;\;8.35 &\;\;\;8.67 \\ 
Total         &   44.19 &   22.81 & 5.61 &   52.84 &    100.00 \\
\midrule
& \multicolumn{5}{c}{Put options}\\
\cmidrule(r){2-6}
Moneyness & (0, 9] & [10, 26] & [27, 33] & >33 & Subtotal\\
\cmidrule(r){1-6}
$<-0.6)$      &\;\;0.00 &\;\;0.00 & 0.00 &\;\;0.17 &\;\;\;0.18 \\ 
$[-0.6,-0.2)$ &\;\;0.38 &\;\;1.38 & 0.48 &\;\;8.34 &   \;10.57 \\ 
$[-0.2,0.2]$  &   20.08 &   19.36 & 3.74 &   27.44 &   \;70.62 \\ 
$(0.2,0.6]$   &\;\;1.09 &\;\;1.76 & 0.37 &\;\;4.06 &\;\;\;7.27 \\ 
$>0.6$        &\;\;0.46 &\;\;0.48 & 0.09 &   10.32 &   \;11.36 \\ 
Total         &   22.01 &   22.98 & 4.67 &   50.34 &    100.00 \\
\bottomrule
\end{tabular}
\renewcommand{\baselinestretch}{0.8}\footnotesize
\begin{minipage}{0.75\textwidth}
This table presents summary statistics for the transaction value of BTC options. The value is measured in USD, i.e., value = volume $\times$ option price (USD) summed in each category. 
The data spans from July 1, 2017, to December 17, 2022. The columns are categorized based on the time to maturity in days. Within each moneyness and maturity category, the entries provide the value proportions in percentage.
\end{minipage}
\end{threeparttable}
\end{center}

\subsection{Estimation of the BVIX}
\label{app:BVIX}
\begin{figure}[ht] 
\begin{center}
    \includegraphics[scale=0.3]{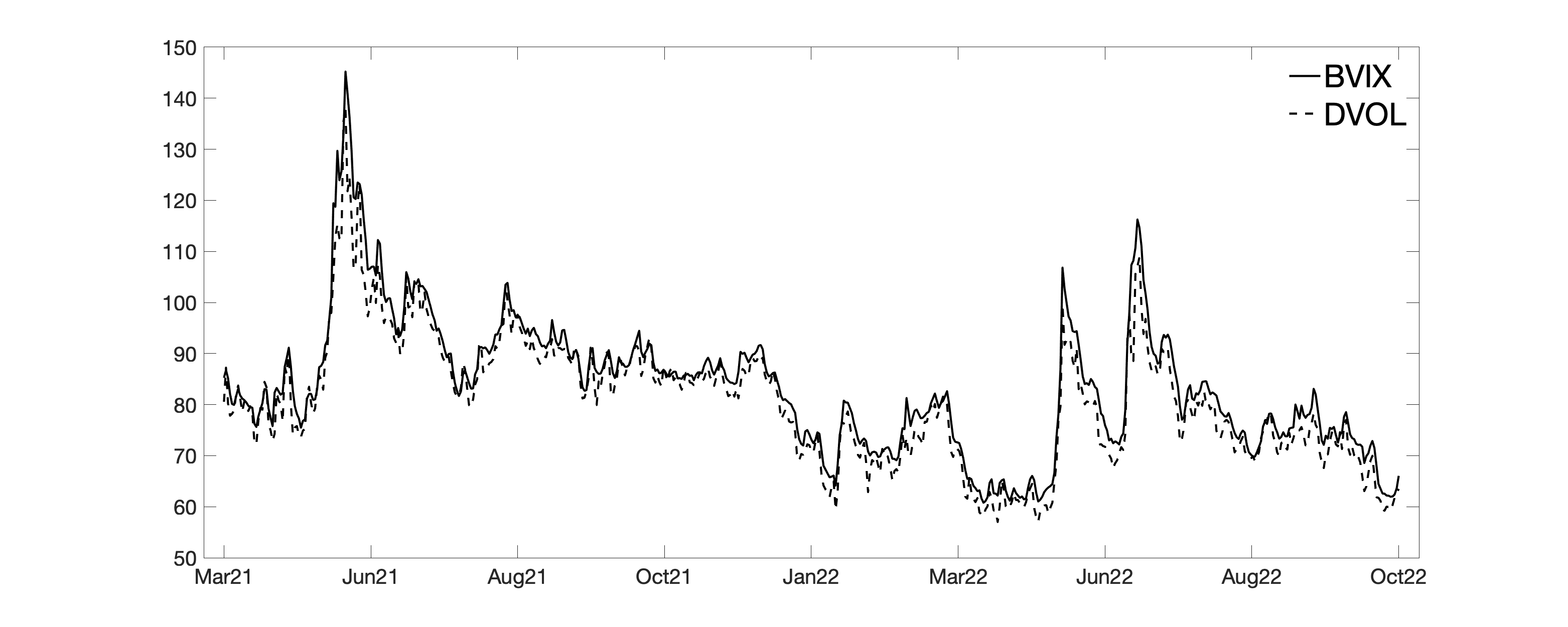} 
\caption{BVIX vs. Deribit Dvol Index}
\label{BVIX}
\end{center}
\end{figure}

The BVIX is calculated using the BTC transaction data as described in Section \ref{sec:data}. We calculate the variances $\sigma_1^2$ and $\sigma_2^2$ by closely following the original VIX methodology of CBOE and interpolate the time-weighted average as
\begin{align}\label{eq:BVIX}
\operatorname{BVIX}_{\tau}=100 \times \sqrt{\left\{N_{T_1} \sigma_1^2\left[\frac{N_{T_2}-\frac{\tau}{365}}{N_{T_2}-N_{T_1}}\right]+N_{T_2} \sigma_2^2\left[\frac{\frac{\tau}{365}-N_{T_1}}{N_{T_2}-N_{T_1}}\right]\right\} \times \frac{365}{\tau}},
\end{align}
where $N_{T_1}=\frac{T_1}{365}$ and $N_{T_2}=\frac{T_2}{365}$ is the time to settlement (in \textit{years}) of the near and next-term options, respectively. A comparison of the BVIX to the Dvol index by Deribit is conducted in Figure \ref{BVIX}, which indicates that both indices are closely related. 

\subsection{Interpolation of the IV Surface} \label{app:interpolation_iv}
 The SVI model is widely popularized due to its parametric specification as well as good performance in the interpolation of IVs. Additionally, unlike some research assuming linearity in $\tau$, relaxing this assumptions allows a more flexible representation of the implied volatility surface, providing a better fit to the data.


\begin{figure}[H]
  \centering
  \begin{tabular}{cc}
    \includegraphics[width=0.5\textwidth]{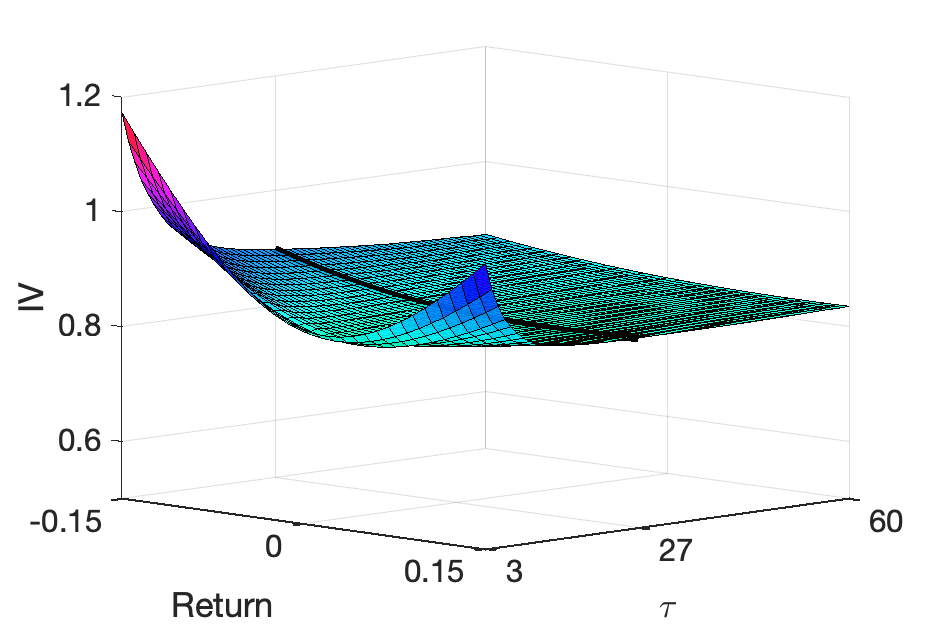} & 
    \includegraphics[width=0.5\textwidth]{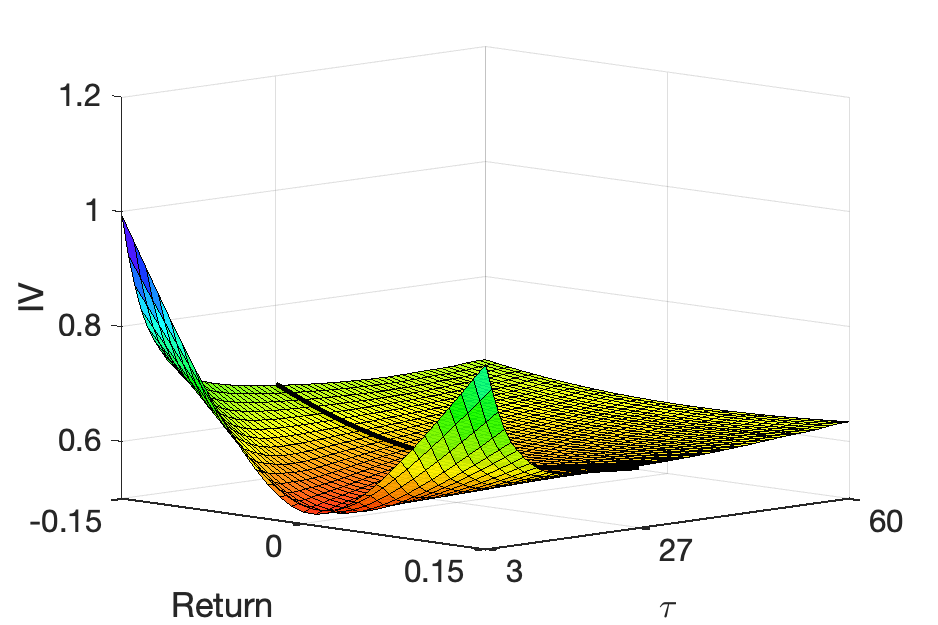} \\
    \\
    \includegraphics[width=0.5\textwidth]{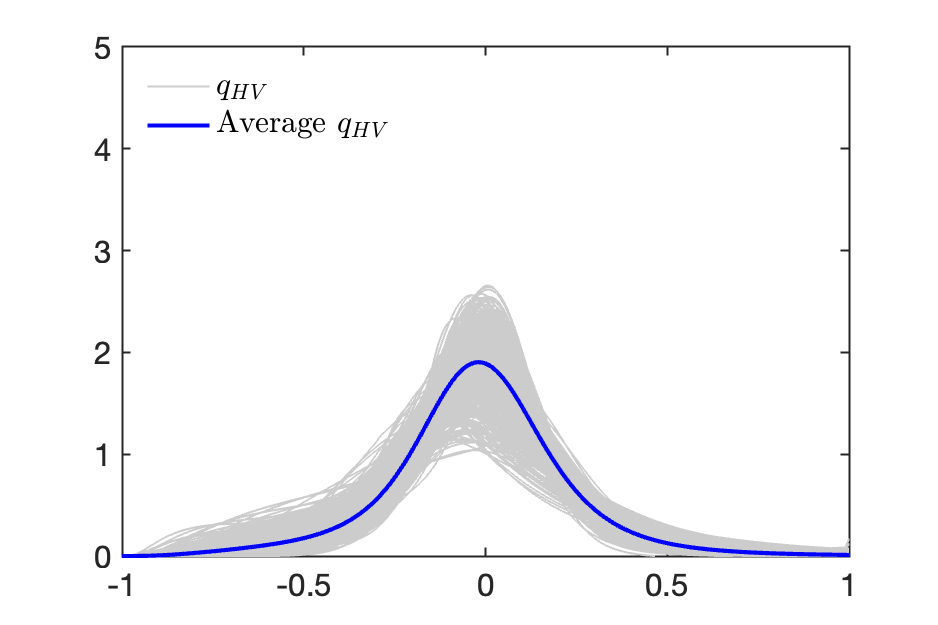} & 
    \includegraphics[width=0.5\textwidth]{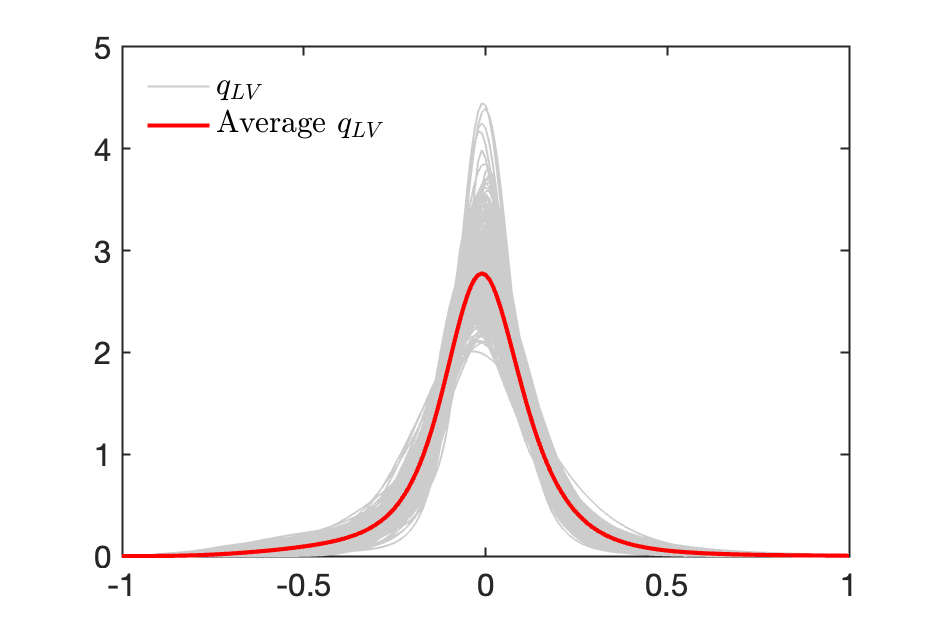} \\
  \end{tabular}
  \caption{First row: Average IV surface interpolated by SVI for the HV cluster (left) and LV cluster (right). The black curve within each panel is the IV for TTM 27 days. Second row: Risk-neutral densities for the HV (left) and LV (right) cluster. The solid curve is the average risk-neutral density for the respective cluster. \label{fig:IV_surface}}  
\end{figure}

\subsection{Bitcoin Premium}

\begin{figure}[H] 
\begin{center}
    \includegraphics[scale=0.6]{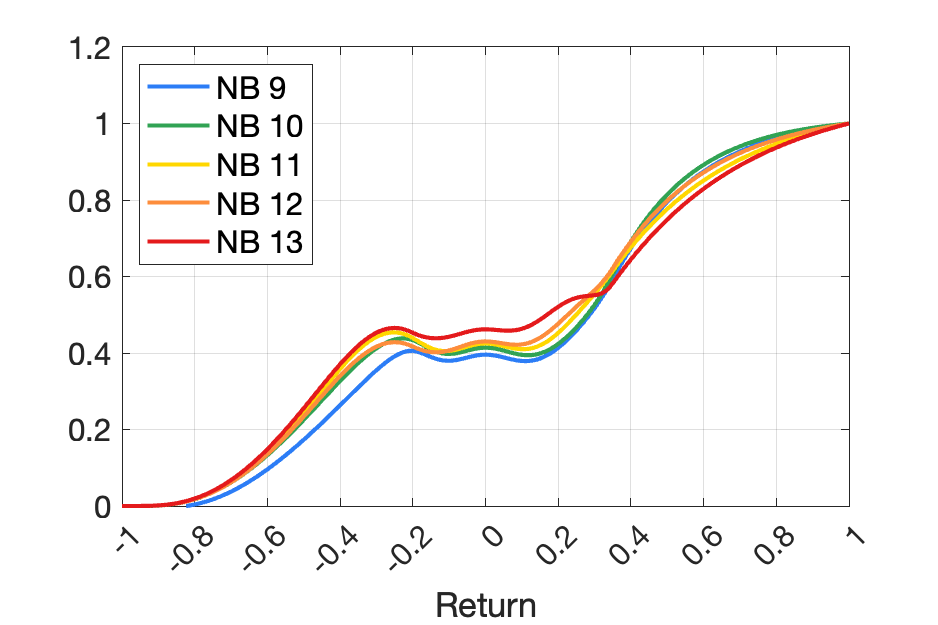} 
\caption{BP across different number of bins (NB) using empirical PDF for the $\p$ density with TTM 27 days.}
\label{fig:app:EP}
\end{center}
\end{figure}

In the estimation of empirical PDF, as $\p$ density, the selection of the smoothing parameter, i.e., the number of equally distant bins, can influence the shape of $\p$. Consequently, this affects the shape of BP. To demonstrate the robustness of BP across various smoothing parameters, Figure \ref{fig:app:EP} shows the robustness of BP with different numbers of bins (NB). In the main text of the paper, we use NB of 11. Despite variations in NB, the basic shape of BP remains relatively consistent.

\subsection{Dimensionality Reduction and Clusters}
\label{app:umap}
\begin{figure}[!htbp]
  \begin{center}
  \subfloat[Principal Component Analysis]{\includegraphics[width=0.5\textwidth]{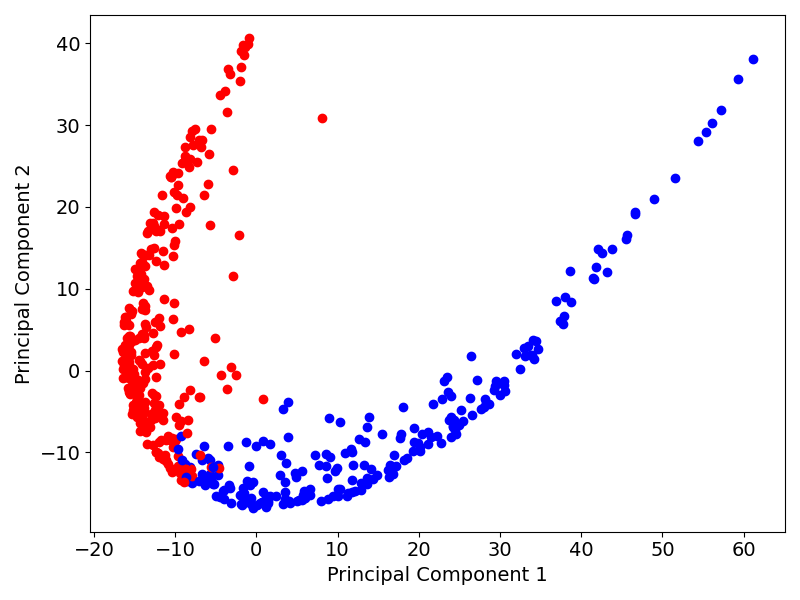}\label{pca}
}
  \subfloat[UMAP]{\includegraphics[width=0.5\textwidth]{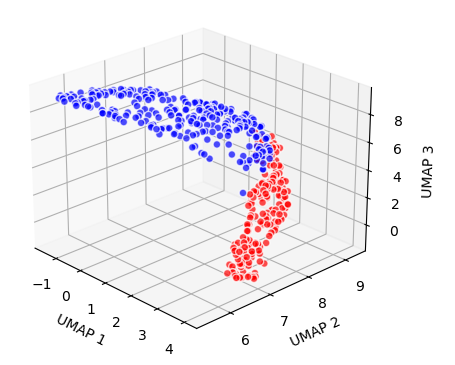}\label{umap}}
  \caption{(a) First two principal components of the Euclidean distance matrix of risk-neutral densities. (b) Three dimensional UMAP of risk-neutral densities. Blue is the HV cluster and red is the LV cluster. Both figures refer to the \textit{multivariate} risk-neutral density composition.}
  \end{center}
\end{figure}

The UMAP (Uniform Manifold Approximation and Projection) is a nonlinear dimensionality reduction technique, recently proposed by \citet{mcinnes2018umap}.
It builds a topological representation of the high dimensional data set and then minimizes the following cross entropy loss function
\begin{align}
\sum_{e \in E} w_h(e) \log \left(\frac{w_h(e)}{w_l(e)}\right)+\left(1-w_h(e)\right) \log \left(\frac{1-w_h(e)}{1-w_l(e)}\right),
\end{align}
where $w_h(e)$ is the weight of the 1-simplex $e$ in the high dimensional case and $w_l(e)$ is the weight of $e$ in the low dimensional case. The set of all possible 1-simplices is represented as $E$. It has been designed to preserve the local as well as the global structure of the data. The result is illustrated in Figure \ref{umap}. Further, we illustrate the first two principal components of the distance matrix in Figure \ref{pca}.

\newpage

\section{Miscellaneous}
This appendix includes miscellaneous results to support our empirical arguments, especially providing further robustness checks. 

\setcounter{figure}{0}
\renewcommand{\thefigure}{B\arabic{figure}}
\setcounter{table}{0}
\renewcommand{\thetable}{B\arabic{table}}

\subsection{Further Cluster Analysis}

\begin{figure}[H]
  \centering
  \begin{tabular}{cc}
    \includegraphics[width=1\textwidth]{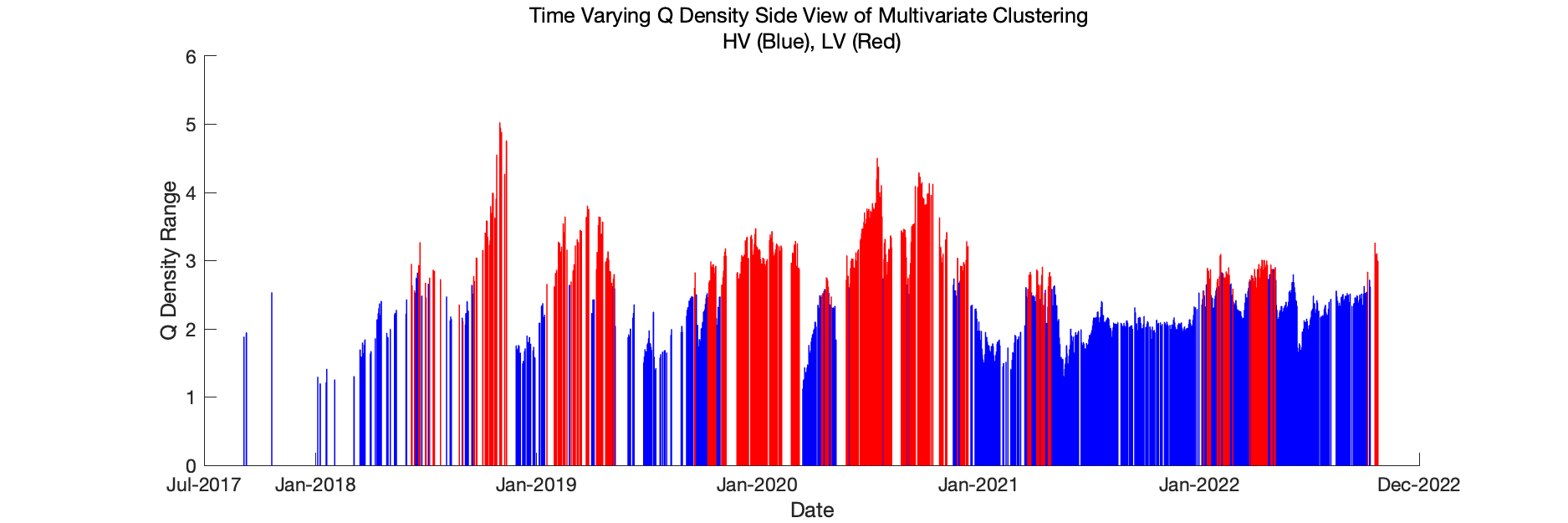} 
  \end{tabular}
  \caption{Time varying risk-neutral densities viewed from the side with time to maturity 27 days. The HV cluster is colored in blue and the LV cluster is colored in red.}\label{fig:side_view_Q_density}
\end{figure}

Figure \ref{fig:side_view_Q_density} provides another way to view $\q$-density in HV and LV clusters for time-to-maturity 27 days. 

\begin{figure}[ht] 
\begin{center}
    \includegraphics[scale=0.5]{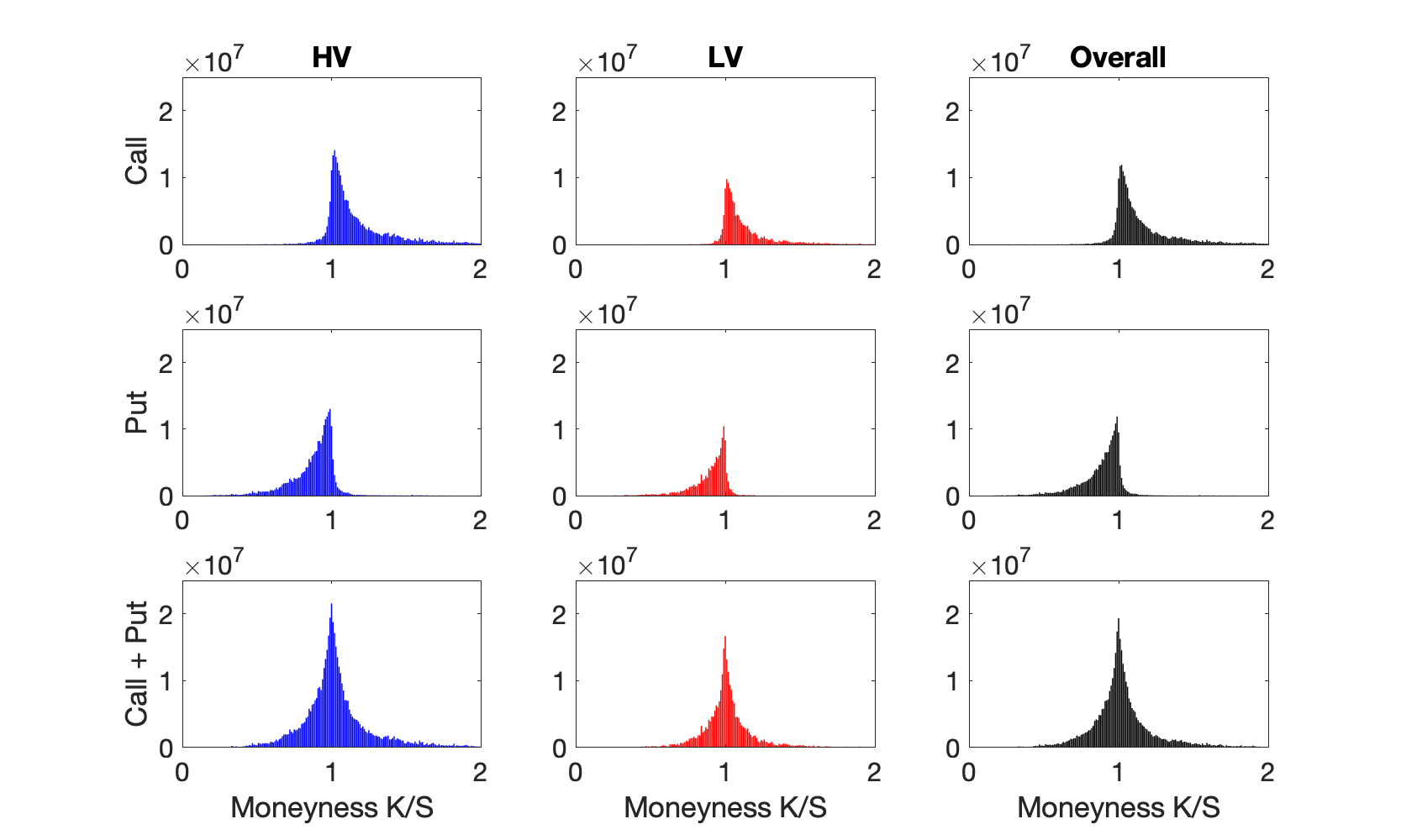} 
\caption{Daily average transaction distribution for different clusters and for call and put, respectively.}
\label{fig:transaction_hist}
\end{center}
\end{figure}

Table \ref{tab:RR_FR} shows the average realized returns (RR) and future returns (FR) on the clustering dates, where returns are simple returns. This indicates that the LV cluster has higher RR, and the HV cluster has higher FR.
Table \ref{tab:logistic_EP_VIX_VRP_ttm27} displays the logistic regression of the clusters on BP, BVIX, and VRP. It reveals that BP calculated by realized return minus risk-free rate can not explain the cluster variation. BVIX shows substantial explainable power, accounting for nearly half of the cluster variance. VRP calculated by $q$ variance minus realized variance is not significant. The coefficients indicate that a higher BVIX index is associated with a higher probability of HV cluster, i.e., the high volatility cluster.
Table \ref{tab:logistic_1to4moment_ttm27} displays the logistic regression examining the relationship between the clusters and the first four moments of risk-neutral density. The variance explains 69\% of the variation in clusters on its own. When combined, all four moments together account for 70\% of the cluster variation. 
Table \ref{tab:logistic_single_factor} presents the logistic regression of clusters with single factors using time-to-maturity 27 days. The dependent variable is the cluster label, while the independent variables include realized returns (RR), realized variance (RV), BVIX, $\q$ variance, VRP(calculated either by BVIX or RV), jumps (including negative and positive jumps) and sentiment index. Significant factors include RR, RV, BVIX, $\q$ variance, and negative and positive jumps separately. Only RV, BVIX, and $\q$ variance have good explanatory power.
Figure \ref{fig:transaction_hist} displays the distribution of daily average BTC option transactions, categorized by clusters and also differentiated into call and put options. This visual illustration confirms that most transactions are OTM for both call and put options, aligning with the summary statistics of transactions we showed in Table \ref{tab:summary_transaction}. Regarding the clustering aspect, the figure indicates that cluster 0 typically experiences higher daily average transactions than cluster 1. 

\begin{center}
\begin{threeparttable}
\centering \footnotesize
\caption{\footnotesize Average realized returns (RR) and future returns (FR) on the clustering dates}\label{tab:RR_FR}
\begin{tabular}{L{0.10\textwidth} C{0.18\textwidth}C{0.18\textwidth}C{0.18\textwidth}}
\toprule
    & Overall & HV & LV \\
\cmidrule(r){2-4}
RR (\%) & 5.64 & 0.41  & 12.05 \\
FR (\%) & 10.27 &  53.83  & -43.09 \\  
Num &  505 & 278  & 227 \\ 
\bottomrule
\end{tabular}
\renewcommand{\baselinestretch}{0.8}\footnotesize
\begin{minipage}{0.73\textwidth}
On each date, we calculate the 27-day realized return (RR) and future return (FR). We report the average RR and FR for each cluster. In comparison, the average 27-day return from Jan 1, 2014 to Dec 31, 2022 is 67.93\%, and the average 27-day return from Jan 1, 2015 to Dec 31, 2022 is 53.34\%. RR and FR are annualized simple returns.
\end{minipage}
\end{threeparttable}
\end{center}

\begin{center}
\begin{threeparttable}
\centering \footnotesize
\caption{\footnotesize Logistic Regression of Clusters on BP, BVIX, and VRP}\label{tab:logistic_EP_VIX_VRP_ttm27}
\begin{tabular}{L{0.1\textwidth} C{0.1\textwidth}C{0.1\textwidth}C{0.1\textwidth}C{0.1\textwidth}C{0.1\textwidth}C{0.1\textwidth}}
\toprule
& \textbf{(1)} & \textbf{(2)} & \textbf{(3)} & \textbf{(4)} & \textbf{(5)}\\
\cmidrule(r){2-7}
Constant       &  -0.20\tmark[{\makebox[0pt][l]{**}}] & -1.21\tmark[{\makebox[0pt][l]{***}}] & -0.25\tmark[{\makebox[0pt][l]{***}}] &  -1.21\tmark[{\makebox[0pt][l]{***}}] &  -0.25\tmark[{\makebox[0pt][l]{***}}]\\ 
               &  (0.09)   & (0.18)   & (0.09)   &  (0.18)   &  (0.09)  \\ 
BP             & 0.12      &          &          &  -0.25    & 0.11\tmark[{\makebox[0pt][l]{***}}]\\ 
               &  (0.09)   &          &          &  (0.18)   &  (0.10)  \\ 
BVIX          &           & -3.75\tmark[{\makebox[0pt][l]{***}}] &          &  -5.37    &          \\ 
               &           & (0.33)   &          &  (0.34)   &          \\ 
VRP            &           &          &  0.15    &           &   0.12   \\ 
               &           &          & (0.11)   &           &  (0.11)  \\ 
\midrule
R2             &    0.00   &    0.49  &   0.00  &    0.50    &    0.01  \\ 
Adj. R2        &   -0.00   &    0.45  &   0.00  &    0.49    &    0.00  \\ 
$T_{HV}$ &  278      &  271     & 271     &  271       &  271     \\ 
$T_{LV}$ &  227      &  211     & 211     &  211       &  211     \\ 
$T$     &  505      &  482     & 482     &  482       &  482     \\ 
\bottomrule
\end{tabular}
\renewcommand{\baselinestretch}{0.8}\footnotesize
\begin{minipage}{0.75\textwidth}
This table displays the logistic regression of the clusters on BP, BVIX, and VRP. The dependent variable is the cluster label. The independent variables are BP, BVIX, and VRP. The number of observations in the HV cluster, LV cluster, and the overall sample is denoted as $T_{HV}$, $T_{LV}$ and $T$, respectively. It reveals that BP calculated by realized return minus risk-free rate can not explain the cluster variation. BVIX shows substantial explainable power, accounting for nearly half of the cluster variance. VRP calculated by $q$ variance minus realized variance is not significant. The coefficients indicate that a higher BVIX index is associated with a higher probability of HV cluster, i.e., the high volatility cluster.
\end{minipage}
\end{threeparttable}
\end{center}

\begin{center}
\begin{threeparttable}
\centering \footnotesize
\caption{\footnotesize Logistic Regression of clusters on first four moments }\label{tab:logistic_1to4moment_ttm27}
\begin{tabular}{L{0.1\textwidth} C{0.1\textwidth}C{0.1\textwidth}C{0.1\textwidth}C{0.1\textwidth}C{0.1\textwidth}C{0.1\textwidth}}
\toprule
& \textbf{(1)} & \textbf{(2)} & \textbf{(3)} & \textbf{(4)} & \textbf{(5)}\\
\cmidrule(r){2-7}
Constant       &  -0.43*** & -2.50*** & -0.46*** &  -0.18    &  -2.19***\\ 
               &  (0.11)   & (0.33)   & (0.11)   &  (0.15)   &  (0.65)  \\ 
Mean           &   1.55*** &          &          &           &   0.30   \\ 
               &  (0.21)   &          &          &           &  (0.40)  \\ 
Variance       &           & -7.70*** &          &           &  -6.28***\\ 
               &           & (0.80)   &          &           &  (1.78)  \\ 
Skewness       &           &          &  1.85*** &           &  -0.04   \\ 
               &           &          & (0.20)   &           &  (0.74)  \\ 
Kurtosis       &           &          &          &   3.93*** &   1.04   \\ 
               &           &          &          &  (0.36)   &  (1.50)  \\ 
\midrule
R2             &    0.14   &    0.69  &   0.24  &    0.60    &    0.70  \\ 
\bottomrule
\end{tabular}
\renewcommand{\baselinestretch}{0.8}\footnotesize
\begin{minipage}{0.75\textwidth}
This table displays the logistic regression examining the relationship between the clusters and the first four moments of risk-neutral density. The dependent variable is the cluster label, and the independent variables include annualized mean, annualized variance, skewness, and excess kurtosis of risk-neutral density. All four moments are standardized. The number of observations in the high-volatility (HV) cluster, low-volatility (LV) cluster, and the overall sample is $T_{HV}=278$, $T_{LV}=227$ and $T=505$, respectively. The variance explains 69\% of the variation in clusters on its own. When combined, all four moments together account for 70\% of the cluster variation.
\end{minipage}
\end{threeparttable}
\end{center}

\begin{sidewaystable}
    \begin{center}
    \begin{threeparttable}
    \centering \footnotesize
    \caption{\footnotesize  Logistic regression of clusters, with single factors TTM=27}\label{tab:logistic_single_factor}
    \begin{tabular}{L{0.10\textwidth} C{0.07\textwidth}C{0.07\textwidth}C{0.07\textwidth}C{0.07\textwidth}C{0.07\textwidth}C{0.07\textwidth}C{0.07\textwidth}C{0.07\textwidth}C{0.07\textwidth}C{0.08\textwidth}}
\toprule
& \textbf{(1)} & \textbf{(2)} & \textbf{(3)} & \textbf{(4)} & \textbf{(5)} & \textbf{(6)} & \textbf{(7)} & \textbf{(8)} & \textbf{(9)} & \textbf{(10)} \\
\cmidrule(r){2-11}
Constant & -0.20** & -1.08*** & -1.22*** & -2.50*** & -0.25*** & -0.25*** &  -0.20** & -0.24*** & -0.23**  &  -0.20***  \\ 
         & (0.09) & (0.17) &  (0.181)  &  (0.33)   &  (0.09)   &  (0.09)   &  (0.09)   &  (0.09)   &  (0.09)    &  (0.09)    \\ 
RR       & 0.15* &      &    &       &       &      &       &       &        &        \\ 
         & (0.09) &     &    &     &     &     &     &     &      &      \\ 
RV       &          & -3.90*** &  &           &           &  &           &  &   &        \\ 
         &          & (0.41) &    &           &           &    &           &    &      &         \\ 
BVIX    &          &  & -3.75*** &           &           &  &  &           &           &      \\ 
         &          &     & (0.33) &           &           &   &   &           &       &    \\ 
$\q$ variance &          &  &          & -7.70*** &           & & &           &           & \\ 
         &          &     &          &           &           &  &     &           &       &      \\ 
VRP(BVIX) &          &           &          &       & 0.15 &           &   &  & &  \\ 
         &          &           &          &     & (0.11) &           &  &     &   &    \\ 
VRP(RV)  &          &           &          &     &           & 0.15 &   & &    &    \\ 
         &          &           &          &    &           & (0.11) &  &  &   &    \\ 
Jump     &          &           &          &           & &           & 0.07 &           &            &   \\ 
         &          &           &          &           &     &           & (0.45) &           &            & \\ 
Jump(-)  &          &           &          &           &   &           &           & 0.60*** &  &      \\ 
         &          &           &          &           &   &           &           & (0.18) &    &    \\ 
Jump(+)  &          &           &          &           &  &           &           &           & -0.48*** &     \\ 
         &          &           &          &           &     &           &           &           & (0.16) &     \\ 
Sentix   &          &  &          &           &           & & &           &           & 0.02 \\ 
         &          &     &          &           &           &     &  &           &   & (0.09) \\ 
\midrule
R2             &    0.00 &    0.28 &   0.49 &    0.69 &    0.00 &    0.00  &    0.00 &    0.02 &    0.02  &    0.00  \\ 
Num. Cluster 0 &  278    &  271    & 271    &  278    &  271    &  271     &  278    &  278    &  278      &  278      \\
Num. Cluster 1 &  227    &  211    & 211    &  227    &  211    &  211     &  227    &  227    &  227      &  227      \\ 
Num. Total     &  505    &  482    & 482    &  505    &  482    &  482     &  505    &  505    &  505      &  505      \\ 
\bottomrule
\end{tabular}
\end{threeparttable}
\end{center}
\end{sidewaystable}

\subsection{Robustness: VRP}
\begin{figure}[!htbp] 
    \begin{center}
    \subfloat[VRP = Q density - RV]{\includegraphics[width=0.5\textwidth]{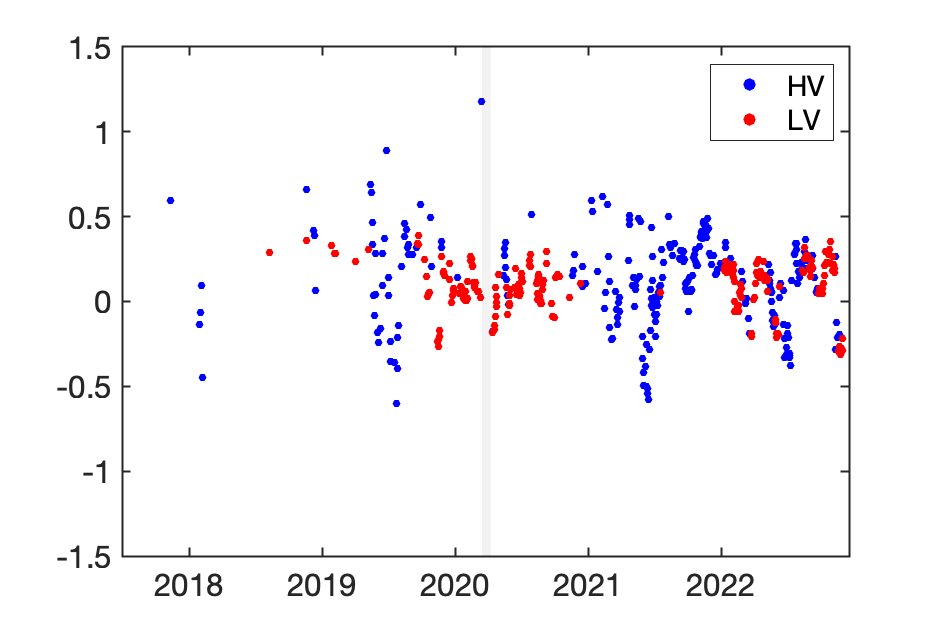}}
    \subfloat[VRP = $\text{BVIX}^2$ - RV]{\includegraphics[width=0.5\textwidth]{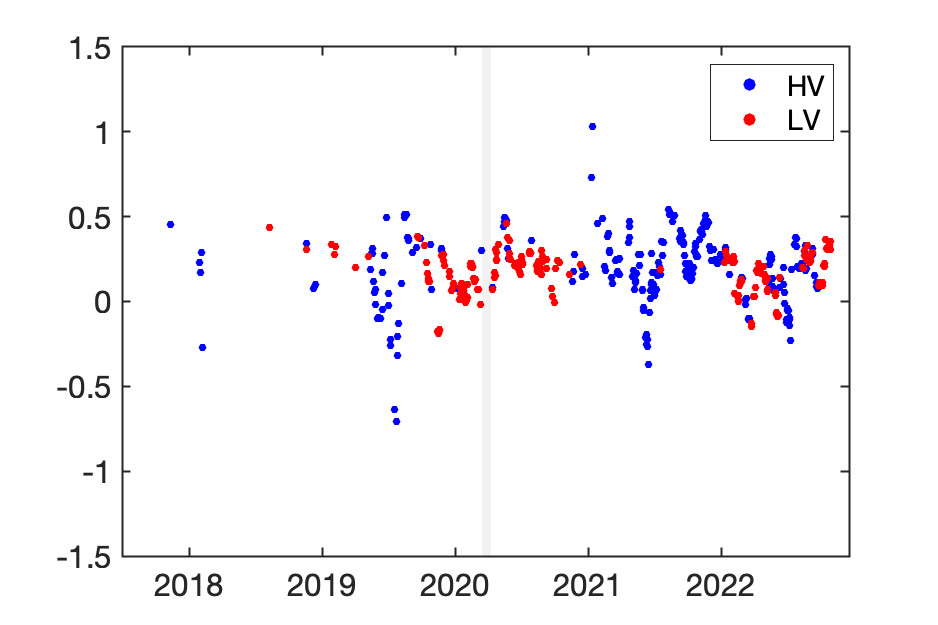}}
    \caption{VRPs over time. In a) VRP is calculated by the second moment of empirical risk-neutral density minus RV. In b) VRP is calculated by $\text{BVIX}^2$ minus RV.}
    \label{fig:app:VRP}
    \end{center}
\end{figure}

To provide more comprehensive VRPs, we provide alternatives of $\q$-variance and $\p$-variance.  For the robustness check of $\p$-variance, we use the second moment of $\p$ density estimated by empirical PDF and present VRP in Table \ref{tab:apx_VRP_RV_2cluster}. Figure \ref{fig:app:VRP} presents two measures of VRP over time; the left is based on the empirical risk-neutral variance, and the right is based on BVIX.

\begin{center}
\begin{threeparttable}
\centering \footnotesize
\caption{\footnotesize Robustness check: Risk Premia}\label{tab:apx_VRP_RV_2cluster}
\begin{tabular}{L{0.15\textwidth} C{0.1\textwidth}C{0.10\textwidth}C{0.1\textwidth}C{0.10\textwidth}C{0.1\textwidth}C{0.10\textwidth}}
\toprule
& \multicolumn{3}{c}{Based on $\q$ density} & \multicolumn{3}{c}{Based on BVIX} \\
\cmidrule(r){2-4}\cmidrule(r){5-7}
                          & Overall & HV  & LV & Overall  & HV  & LV \\
\cmidrule(r){2-4}\cmidrule(r){5-7}
$\Var_{\q}(R)$   &  0.63 &  0.80\tmark[{\makebox[0pt][l]{***}}]  &  0.43\tmark[{\makebox[0pt][l]{***}}] & 0.71 &  0.88\tmark[{\makebox[0pt][l]{***}}]  &  0.50\tmark[{\makebox[0pt][l]{***}}] \\ 
$\Var_{\p}(R)$   &  0.68 &  0.71  &  0.32 & 0.68  &  0.71  &  0.32 \\ 
VRP   &  \textbf{-0.04} & \textbf{0.09\tmark[{\makebox[0pt][l]{***}}]}  &  \textbf{0.11\tmark[{\makebox[0pt][l]{***}}]} & \textbf{0.04} & \textbf{0.16\tmark[{\makebox[0pt][l]{***}}]}  & \textbf{0.18\tmark[{\makebox[0pt][l]{***}}]} \\ 
Observations     &  505 &  278      &  227  & 482      &  271      &  211  \\ 
\bottomrule
\end{tabular}
\renewcommand{\baselinestretch}{0.8}\footnotesize
\begin{minipage}{0.93\textwidth}
$\Var_{\p}(R)$ is $\sigma_p^2$.
\end{minipage}
\end{threeparttable}
\end{center}

\subsection{Cost of Carry}

\begin{figure}[H] 
\begin{center}
    \includegraphics[scale=0.5]{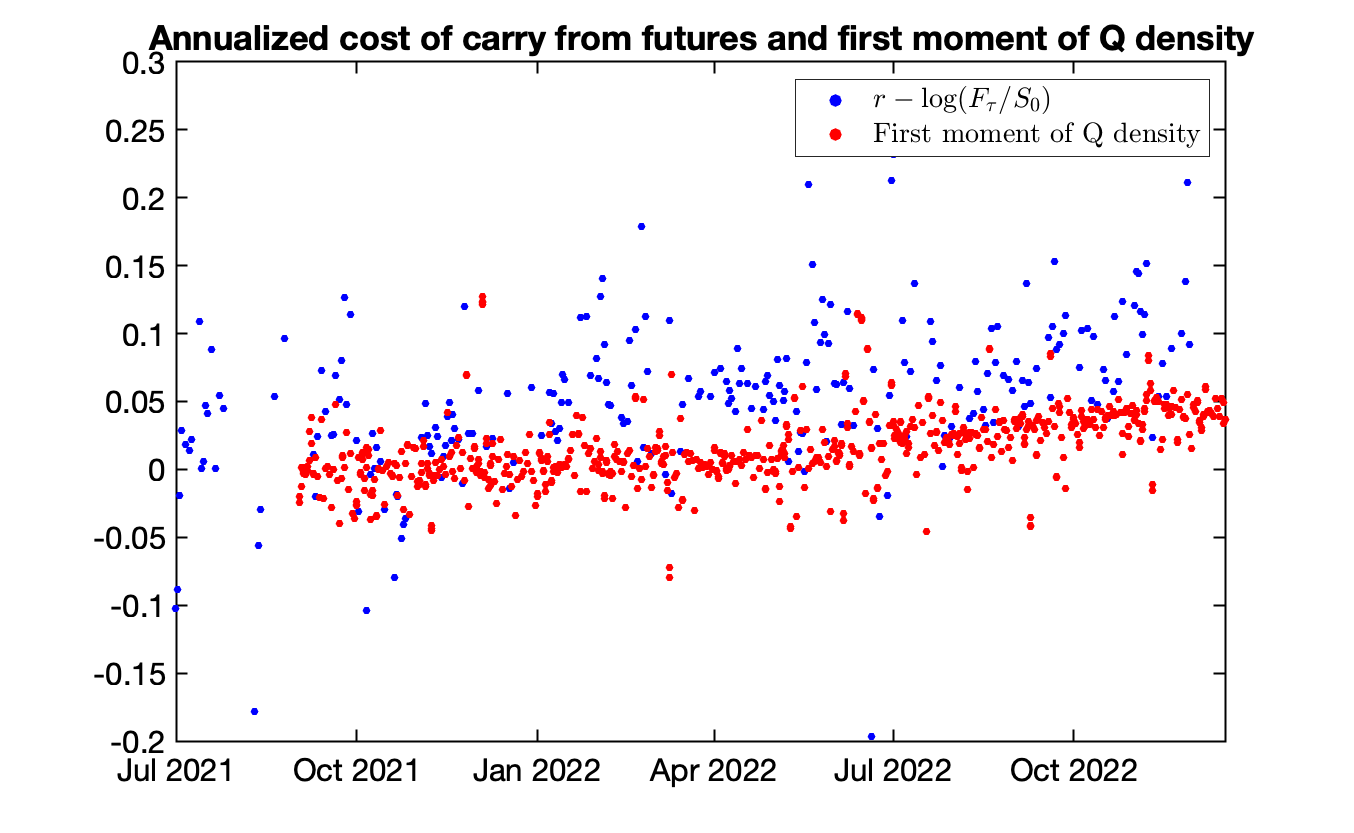} 
\caption{Cost of carry, calculated using BTC futures data from \citet{Liu2023-hy}, compared with the first moment of our estimated $\q$ density}
\label{fig:cost_of_carry_vs_first_Q_moment}
\end{center}
\end{figure}

Figure \ref{fig:cost_of_carry_vs_first_Q_moment} shows the cost of carry, calculated using BTC futures data from \citet{Liu2023-hy}, compared with the first moment of our estimated $\q$ density.

\subsection{Bitcoin comparison with Equity, Bond, and Commodity Markets} \label{sec:assets}

Table \ref{tab:SR} compares the SR of BTC with equity, bond, and commodity markets. It shows the SRs calculated by simple returns and log returns. 
Since BTC returns are asymmetric with high volatility, simple and log returns show different SRs. For simple returns, the SR of BTC is higher than that of other markets, while from log returns, it is around the same level as that of S\&P 500 market.

\begin{center}
\begin{threeparttable}
\centering \footnotesize
\caption{\footnotesize Sharpe Ratio (monthly and annual)}\label{tab:SR}
\begin{tabular}{L{0.15\textwidth} C{0.08\textwidth}C{0.08\textwidth}C{0.08\textwidth}C{0.08\textwidth}C{0.1\textwidth}C{0.08\textwidth}C{0.08\textwidth}}
\toprule
\multicolumn{3}{l}{Panel A: log returns}\\
  & \multicolumn{3}{c}{Monthly} & \multicolumn{3}{c}{Annual} &  \\
\cmidrule(r){2-4}\cmidrule(r){5-7}
& $\hat{\mu}$ (\%) & $\hat{\sigma}$ (\%) & $\widehat{SR}$ & $T\hat{\mu}$ (\%) & $\sqrt{T}\hat{\sigma}$ (\%) & $\sqrt{T}\widehat{SR}$ & Obs. \\
\cmidrule(r){2-8}
BTC & 2.89 & 21.83 &  0.13 & 34.68 & 75.61 &  0.46 & 108 \\
S\&P 500 & 0.68 &  4.41 &  0.15 &  8.14 & 15.28 &  0.53 & 108 \\
Russel 2000 & 0.39 &  5.81 &  0.07 &  4.64 & 20.12 &  0.23 & 108 \\ 
US Bond & 0.05 &  1.10 &  0.05 &  0.61 &  3.80 &  0.16 & 103 \\
US Commodity & -0.06 &  7.06 & -0.01 & -0.73 & 24.45 & -0.03 & 103 \\
\toprule
\multicolumn{3}{l}{Panel B: simple returns}\\
  & \multicolumn{3}{c}{Monthly} & \multicolumn{3}{c}{Annual} &  \\
\cmidrule(r){2-4}\cmidrule(r){5-7}
& $\hat{\mu}$ (\%) & $\hat{\sigma}$ (\%) & $\widehat{SR}$ & $T\hat{\mu}$ (\%) & $\sqrt{T}\hat{\sigma}$ (\%) & $\sqrt{T}\widehat{SR}$ & Obs. \\
\cmidrule(r){2-8}
BTC & 5.40 & 23.33 &  0.23 & 64.81 & 80.82 &  0.80 & 108 \\
S\&P 500 & 0.78 &  4.40 &  0.18 &  9.32 & 15.23 &  0.61 & 108 \\
Russel 2000 & 0.55 &  5.73 &  0.10 &  6.63 & 19.86 &  0.33 & 108 \\ 
US Bond & 0.06 &  1.10 &  0.05 &  0.68 &  3.80 &  0.18 & 103 \\
US Commodity & 0.18 &  6.83 &  0.03 &  2.16 & 23.67 & 0.09 & 103 \\
\bottomrule
\end{tabular}
\renewcommand{\baselinestretch}{0.8}\footnotesize
\begin{minipage}{0.94\textwidth}
{Sharpe Ratio: $SR=(\mu-R_f)/\sigma=(E_P(R)-R_f)/\sqrt{Var_P(R)}$. We use $\hat{\mu}=\frac 1T\sum_{t=1}^T R_t$, $\hat{\sigma}=\sqrt{\frac1T\sum_{t=1}^T (R_t-\hat{\mu})^2}$ to estimate Sharpe Ratio, $\widehat{SR}=(\hat{\mu}-R_f)/\hat{\sigma}$. $R_t$ are \textbf{monthly} log returns in Panel A and simple returns in Panel B. Risk-free rate $R_f=0$. Annualized Sharpe Ratio $\sqrt{T}\widehat{SR}$, with $T$-observations annually, $T=12$ for all. These time series are from January 1, 2014 to December 31, 2022, consistent with the BTC daily prices we used in this paper. For BTC, the annualized simple return $T\hat{\mu}$ (0.64) and volatility $\sqrt{T}\hat{\sigma}$ (0.81) are consistent with the unconditional return $\widehat{\mu}_{\p}$ (0.67) and variance $\widehat{\sigma}^2_\p$ (0.57) in Table \ref{tab:RP_2cluster}, the minor difference might come from the bandwidth we use to calculate $P$-density.
US Bond and US Commodity are represented by the S\&P US Treasury Bond Index and S\&P GSCI Index, respectively, which have been freely available only since 2014-06-01, so they have fewer observations. The Sharpe Ratio calculation refers to \citet{lo2002statistics}. Compared to \citet{Chen2021-vx}, they get a BTC SR of 0.6.
}
\end{minipage}
\end{threeparttable}
\end{center}

\begin{center}
\begin{threeparttable}
\centering \footnotesize
\caption{\footnotesize Correlation matrix}\label{tab:correlation} 
\begin{tabular}{L{0.18\textwidth} C{0.08\textwidth}C{0.08\textwidth}C{0.15\textwidth}C{0.1\textwidth}C{0.18\textwidth}}
\toprule
  & BTC & S\&P 500 & Russel 2000 & US Bond & Global Commodity \\
\cmidrule(r){2-6}
BTC & & -0.02 & -0.01 & 0.00 & 0.06\tmark[{\makebox[0pt][l]{***}}] \\
S\&P 500 & & & 0.88\tmark[{\makebox[0pt][l]{***}}] & -0.22\tmark[{\makebox[0pt][l]{***}}] & 0.32\tmark[{\makebox[0pt][l]{***}}] \\
Russel 2000 & & & & -0.20\tmark[{\makebox[0pt][l]{***}}] & 0.33\tmark[{\makebox[0pt][l]{***}}] \\ 
US Bond & & & & & -0.17\tmark[{\makebox[0pt][l]{***}}] \\
Global Commodity & \\
\bottomrule
\end{tabular}
\renewcommand{\baselinestretch}{0.8}\footnotesize
\begin{minipage}{0.92\textwidth}
{For equity markets, we use S\&P 500 and Russel 2000 indices. For the bond market, we utilize the S\&P US Treasury Bond Index. For Global commodities, we use the S\&P GSCI Index. Correlation is performed using a t-test ($H_0:$ no correlation). The t-statistic is calculated as $t=\frac{corr\sqrt{n-2}}{\sqrt{1-corr^2}}$, where $corr$ represents the correlation coefficient and $n$ is the sample size. Significance levels are denoted by $1\%(^{***})$, $5\%(^{**})$ and $10\%(^{*})$. These time series are from June 6, 2014 to December 31, 2023.
}
\end{minipage}
\end{threeparttable}
\end{center}

Table \ref{tab:correlation} presents the correlation matrix among Bitcoin, S\&P 500, Russel 2000, US Bond, and Global Commodity indices. This indicates that BTC has no significant correlation with equity and bond markets but is more correlated with commodity markets. 

\subsection{Related literature on EP, VRP, and BTC options}\label{sec:premia}

\begin{center}
\begin{threeparttable}
\centering \footnotesize
\caption{\footnotesize Stock Equity premium and Bitcoin premium}\label{tab:literature_EP_BP}
\begin{tabular}{L{0.40\textwidth} C{0.26\textwidth} C{0.26\textwidth}}
\toprule
\multicolumn{3}{l}{Panel A: Stock}\\
 & Equity premium & Time \\
\cmidrule(r){2-3}
\citet{heston2023pricing}   &  0 - 8.32\% &  1996 - 2019\\ 
\citet{Tetlock2023-ry}   &  8.64\% &  1996 - 2021\\ 
\citet{chabi-yo2023decomposition}   &  8.72\% &  1996 - 2019\\ 
\cmidrule(r){1-3}
\multicolumn{3}{l}{Panel B: BTC}\\
 & BTC premium & Time \\
\cmidrule(r){2-3}
\citet{Chen2021-vx}   &  48.12\%  &  Feb 2018 - Sep 2020\\ 
\citet{Foley2022-py}   &  80\%  &  2018 - 2020\\ 
\citet{Wilson2024-cz}   &  273.6\% &  Apr 2010 - Feb 2023\\ 
   &  52.68\%  &  Dec 2013 - Feb 2023\\ 
   &  75\%  &  May 2017 - Feb 2023\\ 
\bottomrule
\end{tabular}
\renewcommand{\baselinestretch}{0.8}\footnotesize
\end{threeparttable}
\end{center}

\begin{center}
\begin{threeparttable}
\centering \footnotesize
\caption{\footnotesize Stock and Bitcoin Variance Risk premium}\label{tab:literature_BVRP}
\begin{tabular}{L{0.36\textwidth} C{0.10\textwidth} C{0.25\textwidth} C{0.2\textwidth}}
\toprule
\multicolumn{3}{l}{Panel A: Stock}\\
 & VRP & Time & Definition \\
\cmidrule(r){2-4}
\citet{Bakshi2003-nb}   &  (-) &  Jan 1, 1988 - Dec 31, 1995 & reg. coef. b/w DH gains and vega \\
\citet{Carr2009-oe}   &  -2.74\% &  Jan 1996 - Feb 2003 & $\p-\q$\\
\citet{Bollerslev2009-sb}   &  18.30$\%^2$ monthly &  1990 - 2007 & $\q-\p$\\
\citet{Todorov2010-cy}   &  -0.4015$\%^2$ daily &  1990 - 2002 & $\p-\q$\\
\citet{Bekaert2014-la}   &  (+) &  Jan 2, 1990 - Oct 1, 2010 & $\q-\p$\\
\citet{zhou2018variance}   &  (+) &  1990 - 2015 & $\q-\p$\\
\citet{rombouts2020dynamics}   &  17\% &  Jan 1990 - Sep 2015 & $\q-\p$\\
\citet{heston2023pricing}   &  -6.06\% - 0 &  1996 - 2019 & $\lambda v_t$\\
\citet{Tetlock2023-ry}   &  1.56\% &  1996 - 2021 & $\q-\p$\\
\cmidrule(r){1-4}
\multicolumn{3}{l}{Panel B: BTC}\\
 & VRP & Time & Definition \\
\cmidrule(r){2-4}
\citet{Alexander2021-ai}   &  mostly (-) &  Mar 2019 - Mar 2020 & $\p-\q$\\
\bottomrule
\end{tabular}
\renewcommand{\baselinestretch}{0.8}\footnotesize
\begin{minipage}{0.98\textwidth}
{In \citet{Bollerslev2009-sb}, the VRP of S\&P 500 index is reported as 18.30$\%^2$, so the annualized VRP is approximately
 $18.30\%^2 \times 12 = 219.6\%^2$ or $2.196\%$. In \citet{Todorov2010-cy}, the VRP is reported as -0.4015 in variance unit ($\%^2$), therefore the annualized VRP is 
 $-0.4015\%^2 \times 252 = 101.178\%^2$ or $1.012\%$. }
\end{minipage}
\end{threeparttable}
\end{center}

Table \ref{tab:literature_EP_BP} summarizes empirical results from the literature about the stock equity premium and the Bitcoin premium. Notably, \citet{chabi-yo2023decomposition} decomposed EP into bad, moderate, and good states by moneyness. They find that the average contribution in bad states is about 20\% of EP, while in crisis, this number increases to 60\%. In a stable period, the central state contributes 80\%. They claim they also decompose higher moments of VRP in online supplements. Another point to notice is that \citet{Wilson2024-cz} defines Bitcoin Premium as excess returns of BTC returns minus stock returns. Further, table \ref{tab:literature_BVRP} reports results on the literature of equity VRP and Bitcoin VRP. \citet{Bollerslev2009-sb} and the best model of \citet{Bekaert2014-la} have similar tendency, as shown in \citet{Cheng2019-yj}.

\printbibliography
\end{refsection}

\end{document}